\documentclass[11pt,british]{article}
\usepackage[utf8]{inputenc}
\usepackage{mathtools}
\usepackage{amsmath}
\usepackage{amsthm}
\usepackage{amssymb}
\usepackage{setspace}
\usepackage{microtype}
\usepackage{babel}
\usepackage{stmaryrd}
\usepackage{slashed}
\usepackage{cancel}
\usepackage{bbm}
\usepackage{booktabs}
\usepackage{bm}
\usepackage{float}
\usepackage[final]{showlabels}

\makeatletter
\usepackage{tikz}
\usepackage{tikz-cd}
\usetikzlibrary{backgrounds}
\usepackage[framemethod=tikz]{mdframed}
\AtBeginEnvironment{mdframed}{%
\tikzset{every picture/.style={}}%
}
\mdfsetup{roundcorner=.5ex}

\theoremstyle{definition}
\newtheorem*{defn*}{Definition}

\usepackage{jheppub}                   
\usepackage[linecolor=blue,backgroundcolor=blue!25,bordercolor=blue,textsize=scriptsize]{todonotes}

\makeatletter
\gdef\@fpheader{\ }                    
\makeatother

\usepackage{setspace}
\usepackage{slashed}	 	
\usepackage{amsfonts} 		
\SetTracking{encoding={*}, shape=sc}{0} 	
\usepackage{color} 		
\definecolor{darkblue}{rgb}{0.0,0.0,0.3} 	
\allowdisplaybreaks		
\date{\today} 		
\numberwithin{equation}{section}	

\makeatletter
\g@addto@macro\bfseries{\boldmath}
\makeatother

\let\originalleft\left
\let\originalright\right
\renewcommand{\left}{\mathopen{}\mathclose\bgroup\originalleft}
\renewcommand{\right}{\aftergroup\egroup\originalright}

\theoremstyle{theorem}
\newtheorem{theorem}{Theorem}

\newcommand{\Mcal}{\mathcal{M}}
\newcommand{\Ncal}{\mathcal{N}}
\newcommand{\Xcal}{\mathcal{X}}
\newcommand{\Wcal}{\mathcal{W}}

\newcommand{\Ical}{\mathcal{I}}

\newcommand{\Ccal}{\mathcal{C}}
\newcommand{\Ycal}{\mathcal{Y}}
\newcommand{\Scal}{\mathcal{S}}
\newcommand{\Hcal}{\mathcal{H}}

\newcommand{\bbR}{\mathbb{R}}
\newcommand{\bbC}{\mathbb{C}}
\newcommand{\bbZ}{\mathbb{Z}}

\newcommand{\dd}{\mathrm{d}}
\newcommand{\ee}{\mathrm{e}}
\newcommand{\co}{\mathrm{co}}
\newcommand{\ex}{\mathrm{ex}}

\newcommand{\wrap}{\mathfrak{w}}
\newcommand{\winding}{w}

\newcommand{\Spin}{\mathrm{Spin}}

  \definecolor{cambridgeblue}{rgb}{0.64, 0.76, 0.68}
	\definecolor{lapislazuli}{rgb}{0.15, 0.38, 0.61}
\definecolor{awesome}{rgb}{1.0, 0.13, 0.32}
\definecolor{aureolin}{rgb}{0.99, 0.93, 0.0}
\definecolor{almond}{rgb}{0.94, 0.87, 0.8}
\definecolor{antiquewhite}{rgb}{0.98, 0.92, 0.84}

\newcommand{\dr}{\mathrm{d}}
\newcommand{\pd}{\partial}

\def\be{\begin{equation}}
\def\ee{\end{equation}}

\title{Brane wrapping, AKSZ sigma models, and QP manifolds}

\subheader{\hfill\textrm{MI-HET-793}}

\author[a]{Alex S. Arvanitakis}
\emailAdd{alex.s.arvanitakis@vub.be}
\author[b]{and David Tennyson}
\emailAdd{dtennyson@tamu.edu}

\affiliation[a]{Theoretische Natuurkunde, Vrije Universiteit Brussel, and the International Solvay Institutes, Pleinlaan 2, B-1050 Brussels, Belgium}
\affiliation[b]{Mitchell Institute for Fundamental Physics and Astronomy, Texas A\&M University, College Station, TX, 77843, USA}

\abstract{We introduce a technique to realise brane wrapping and double dimensional reduction in the context of AKSZ topological sigma models and also in their target spaces, which are symplectic $L_n$-algebroids (i.e.~QP-manifolds). Our procedure involves a novel coisotropic reduction combined with an AKSZ transgression that realises degree-shifting; the reduced QP-manifold depends on topological data of the `wrapped' cycle. We check our procedure against the known rules for fluxes under wrapping in the context of M-theory/type IIA duality, and we also find a new relation between Courant algebroids and Poisson manifolds.}

\begin{document}

\maketitle

\section{Introduction}
In a series of recent papers \cite{Arvanitakis:2018cyo,Arvanitakis:2021wkt,Arvanitakis:2022fvv} we have been establishing a correspondence between \textbf{(BPS) $p$-branes} in String/M-theory on one hand and \textbf{symplectic $L_{p+1}$-algebroids} on the other hand. The latter can be thought of as appropriate generalisations of the (exact) \emph{Courant algebroid} that encodes the generalised geometry of type II backgounds with Neveu-Schwarz 3-form $H$ (but without Ramond-Ramond fluxes); a Courant algebroid is precisely a symplectic $L_n$-algebroid of degree $n=2$ \cite{roytenberg1999courant}. In this correspondence the exact Courant algebroid (which is classified by the de Rham cohomology class of $H$  \cite{Severa:2017oew}) is associated to the fundamental string (which couples to $H$ electrically via a Wess-Zumino term). For other branes, such as the M2 and M5 branes in M-theory, or even/odd D-branes in type II string theory, the corresponding algebroids are roughly speaking the ones that are classified by whichever fluxes couple \textit{electrically} to the brane in question; such algebroids can be thought of as generalisations of Courant algebroids to e.g.~M-theory scenarios, see \cite{Arvanitakis:2019cxy} for this point of view, and \cite{Sun:2022pgs} for an alternative.

In general, the correspondence is between a ``physical'' $p$-brane (an F1, M2, M5\dots), a symplectic $L_n$-algebroid for $n=p+1$, and a \emph{topological} AKSZ brane sigma model of dimension $p+2$. This can be schematically summarised in the following diagram:
\begin{equation}
\label{diagram}
\begin{tikzcd}
&  \text{symplectic $L_n$-algebroid} \ar[dl,"\text{AKSZ}"] \ar[dr,"\text{brane phase space}"] &  \\
\text{topological $n$-brane} \ar[rr,"\text{boundary condition}",dotted]  & & \text{physical $(n-1)$-brane}
\end{tikzcd}
\end{equation}
Less tersely, the symplectic $L_n$-algebroid --- that is classified by a certain collection of fluxes --- determines a topological $n$-brane sigma model via the AKSZ construction \cite{Alexandrov:1995kv}. When the $n$-brane has an $(n-1)$-brane boundary, an inflow-type argument with an appropriate boundary condition produces the WZ term that couples those same fluxes to the $(n-1)$-brane \cite{Arvanitakis:2018cyo,Arvanitakis:2022fvv}\footnote{A slightly different boundary condition for the AKSZ sigma model can produce the entire $(n-1)$-brane lagrangian, including kinetic terms. This was done for the fundamental string by \v{S}evera \cite{Severa:2016prq}. The other cases have not yet been considered in the literature.}. The algebroid also determines the corresponding $(n-1)$-brane more directly via the \emph{brane phase space} construction that yields the Poisson algebra of brane currents on phase space \cite{Arvanitakis:2021wkt} (i.e.~in the hamiltonian formulation of brane dynamics).

This correspondence between branes and algebroids motivates the question: given that the String/M-theory duality web acts on the branes, \textbf{how is the  duality web realised on the algebroid side?} Heretofore this was only known  for dualities that preserve worldvolume dimension; see \cite{Arvanitakis:2021lwo} for T-duality, and \cite{Arvanitakis:2022fvv} for M-theory/type IIA duality along a transverse M-theory circle. An example of the latter is the emergence of a D2 brane given an M2 brane that does not wrap the M-theory circle, whose algebroid avatar is symplectic reduction modulo the $\mathrm{U}(1)$ action.

In this paper we provide an algebroid realisation for the \textbf{brane wrapping} operation. In the string theory picture, this sends a $p$-brane to the $(p-d)$-brane found by wrapping the original brane around a $d$-dimensional cycle on target space and then shrinking the volume of the cycle to zero. (Since both the dimensionality of the brane and that of the target space are reduced in this way, this is also known as \emph{double dimensional reduction}.) The most basic example is M-theory/IIA duality, where M2 branes wrapped around the compactified 11th dimension give rise to fundamental strings in 10 dimensions \cite{Duff:1987bx}. This already poses a puzzle: the corresponding algebroids are of degree $n=p+1=3$ (for the M2 brane) and $n=2$ (for the F1); \textbf{what is the mathematical operation that accounts for this degree shift?}

The mystery is resolved in the supergeometric formulation of symplectic $L_n$-algebroids, defined by the data of a \emph{QP manifold} $(\Mcal,\omega,Q)$ where $\Mcal$ is a non-negatively graded manifold, $\omega$ a symplectic form of degree $n$, and $Q$ a nilpotent vector field of degree $1$, hamiltonian for $\omega$. Given a compact manifold $X$ of dimension $d$ --- to be identified with the cycle to be `wrapped' --- the odd tangent bundle $\Xcal\equiv T[1]X$ possesses an integration measure $\int_\Xcal:C^\infty(\Xcal)\to \mathbb R$ of degree $-d$, namely the integral of differential forms. Then the mapping space
\be
\Mcal^\Xcal\equiv \operatorname{maps}(\Xcal\to\Mcal)
\ee
possesses a P-structure of degree $(n-d)$, provided by the AKSZ construction. This is the correct degree shift; however, this manifold is infinite dimensional, and its structure sheaf is not non-negatively graded, so it cannot be the sought-after symplectic $L_{n-d}$-algebroid.

\subsubsection*{A `brane wrapping' for QP manifolds.}

We introduce a coisotropic reduction of the space $\Mcal^\Xcal$ to a finite-dimensional QP-manifold that resolves both issues. This resolution is heavily motivated by the intuitive string-theoretic picture of brane wrapping. We deal with the case where the body of $\Mcal$ is a product $N\times X$, seen as a trivial bundle with fibre $X$, and we select a map $N\hookrightarrow \operatorname{maps}(X\to N\times X)$, as in the figure
\begin{figure}[H]
    \centering
    \includegraphics[width=0.44\textwidth]{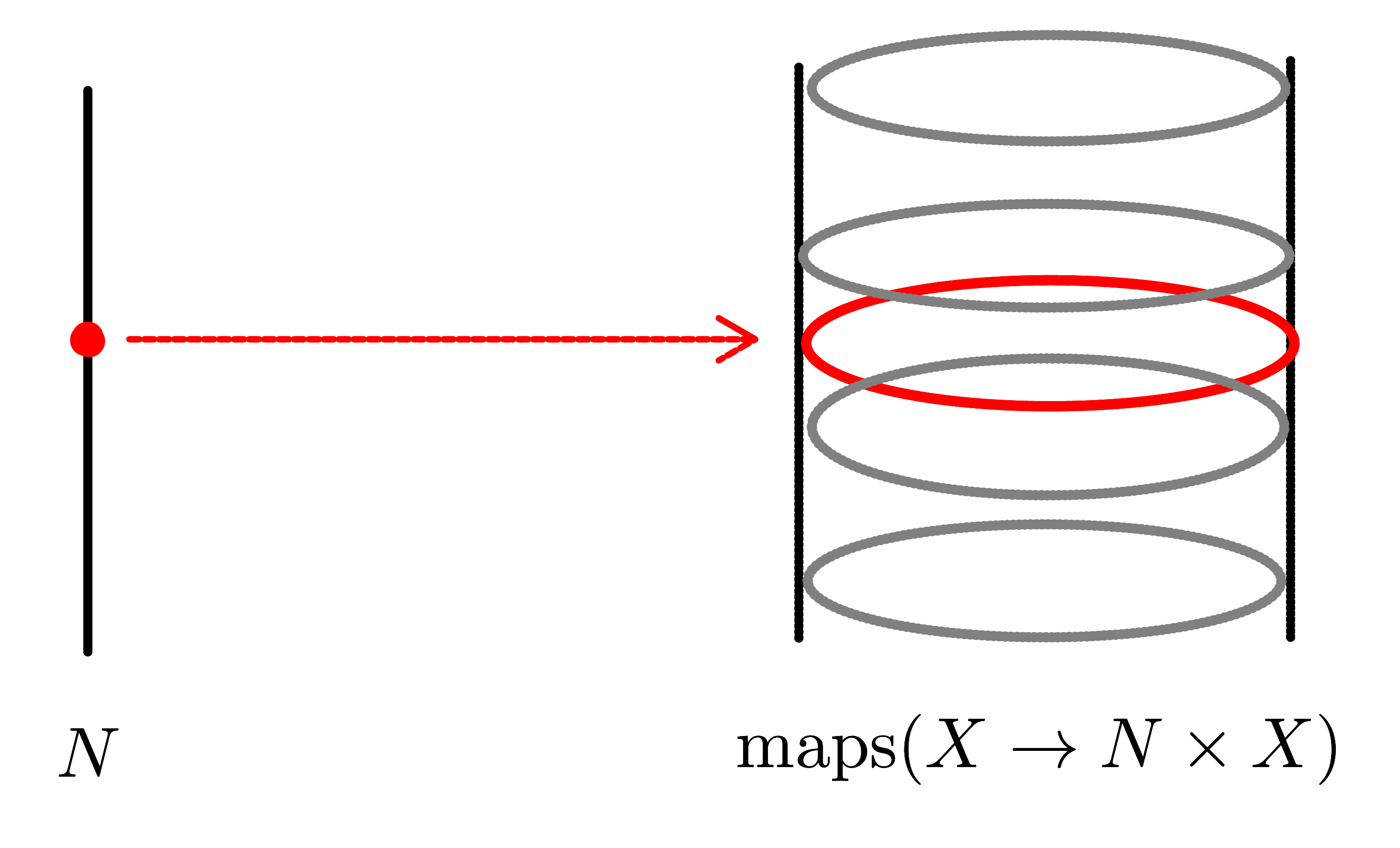}
    \caption{The wrapping map specification, for $N=\mathbb R$, $X=S^1$.}
    \label{fig:my_label}
\end{figure}
\noindent The idea is that each point $n\in N$ is mapped to the cycle of $N\times X$ that shrinks to zero size in the double dimensional reduction procedure. Since $\operatorname{maps}(X\to N\times X)$ is disconnected, with connected components corresponding to different winding sectors (as they would be called in physics), the choice of map $N\hookrightarrow \operatorname{maps}(X \to N\times X)$ includes a choice of winding. On the string theory side, double dimensional reduction indeed depends on winding: for instance, an M2 brane wound $w$ times around the M-theory circle yields a fundamental string coupled to the $H$-flux $w H$. Since the algebroids corresponding to these branes via the diagram \eqref{diagram} are defined by the same fluxes, we expect winding dependence in the obtained algebroid, and we will indeed find it.

In more detail: we start with the data of an NQP --- ``N'' for non-negatively graded --- manifold $\Mcal$ with body $M$ and a `source' Q manifold $\Xcal=T[1]X$ as above, along with a \textbf{wrapping map} $\wrap:X\to M$ that defines a degree-zero submanifold $N\hookrightarrow \operatorname{maps}(X \to M)$. We then produce a \emph{finite-dimensional, non-negatively graded} QP manifold $\Wcal$, whose P-structure has degree $n-d$; we will call $\Wcal$ the \textbf{wrapped algebroid}, and we will call our procedure \textbf{(brane) wrapping}. The wrapping of QP manifolds/symplectic $L_n$ algebroids is then a reduction of $\Mcal^\Xcal$ with respect to a coisotropic submanifold $\Ccal$ which may be thought of as the lift of $N\hookrightarrow \operatorname{maps}(X \to M)$ to a graded submanifold of $\operatorname{maps}(\Xcal \to \Mcal)=\Mcal^\Xcal$. The output QP manifold $\Wcal$ depends on the choice of wrapping map $\wrap$ only up to homotopy.

In fact we were able to generalise beyond the case $M=N\times X$ (that was pictorially outlined above) to the case $M=N\times Y$, with $Y$ and $X$ not necessarily of the same dimension, even; then the wrapping is a map $\wrap:X\to N\times Y$, and $d=\dim X$ controls the degree/dimensionality shifts as before. This generalisation allows us to accommodate at least one example which might be of interest outside of string theory, namely the wrapping of a Courant algebroid into a Poisson manifold discussed in Section \ref{sec:Courant->Poisson}, which has $\dim Y=0$. When  $\dim X=n+1$ in addition to $\dim Y=0$ (so that $\Mcal^\Xcal$ has a degree $-1$ P structure) our wrapping procedure agrees with that of \cite{Bonechi:2010tbl}. Our approach gives a complementary perspective to the Losev-trick based `wrapping'-style reductions of \cite{Kokenyesi:2018ynq,Cattaneo:2009zx}, and to that of \cite{Chatzistavrakidis:2019rpp,Kokenyesi:2018xgj}.

\subsubsection*{Brane wrapping and AKSZ sigma models.}

Our `brane wrapping' reduction --- from a QP manifold $\Mcal$ to a QP manifold $\Wcal$ --- also induces a reduction of the corresponding AKSZ topological field theories. Essentially, the two reductions commute, as in the schematic diagam
\begin{equation}
\begin{tikzcd}[column sep=large]
\Mcal \ar[r,"\text{wrapping}"]\ar[d,"\text{AKSZ}"]  &  \Wcal \ar[d,"\text{AKSZ}"] \\
\Mcal^{\Xcal\times \Scal} \ar[r,dotted]  & \Wcal^\Scal
\end{tikzcd}
\end{equation}
Here $\Mcal^{\Xcal\times \Scal}$ and $\Wcal^\Scal$ are P-manifolds of degree $-1$ created by the AKSZ construction for $\Scal$ of appropriate dimension.
The dotted arrow corresponds to a reduction of $\Mcal^{\Xcal\times \Scal}$ with respect to the coisotropic submanifold $\Ccal^{\Scal}\equiv \operatorname{maps}(\Scal,\Ccal)$, for $\Ccal$ the coisotropic submanifold that appears in the `wrapping' reduction $\Mcal\to \Wcal$. This `dotted' reduction always exists and is compatible with the AKSZ/BV master actions if the wrapping reduction does.

We provide the argument for the reduction of AKSZ sigma models in section \ref{sec:lifting_AKSZ}, along with an example: the reduction of a topological 3-brane sigma model (corresponding to the M2 brane symplectic $L_3$-algebroid) to a Courant sigma model (corresponding to the fundamental string symplectic $L_2$-algebroid). This provides an important consistency check: if we were to derive the corresponding physical brane sigma models, e.g.~by introducing boundaries and using an inflow-type argument as in \cite{Severa:2016prq,Arvanitakis:2018cyo}, we would find that the electric Wess-Zumino flux coupling has the correct winding dependence.

\subsubsection{Structure of the paper.} 

In section \ref{sec:wrapping_QP}, we describe the general procedure for wrapping QP manifolds. We provide the conditions required of the QP structure on $\Mcal$ and define the coisotropic ideal $\Ical \subset C^{\infty}(\Mcal^{\Xcal})$ (that defines the coisotropic submanifold $\Ccal$) in general. We show that it is well-defined and perform the reduction. The next three sections provide a multitude for examples. (If the reader finds the notation of section \ref{sec:wrapping_QP} too terse, they may find it useful to first work their way through the examples before coming back to the general procedure.) Section \ref{sec:dimX=0} covers the case where $\dim X = 0$. In this case, we do not get any wrapping and our reduction is very similar to conventional dimensional reduction \cite{Bonechi:2012kh}. In section \ref{sec:dimY=0} we consider examples where $\dim X \neq 0$, but the wrapping map $\wrap$ is trivial in homotopy. These provide examples which are simple but still present some of the main features of the reduction. Among these is the reduction of a Courant algebroid to a Poisson manifold given in section \ref{sec:Courant->Poisson}. In section \ref{sec:examples_X=Y}, we consider examples relevant for physics and wrap string/M-theory branes on various manifolds. In section \ref{sec:lifting_AKSZ} we show how our procedure naturally lifts to a reduction of the AKSZ theory from $\Mcal^{\Xcal\times \Scal}$ to $\Wcal^{\Scal}$. Section \ref{sec:conclusions} is left for comments and outlook. The appendices cover our notation (appendix \ref{app:notation}), some key properties and conventions of QP manifolds (appendix \ref{app:QP_manifolds}), and a review of coisotropic reduction in the graded context (appendix \ref{app:CoisoreductionOfPoisson}).

\section{Wrapping QP manifolds}\label{sec:wrapping_QP}

We will describe a process of creating new QP manifolds from old, which effectively generalises the notion of dimensional reduction, that we describe as `wrapping' QP manifolds. The nomenclature arises due to the consistency of this process with the AKSZ construction \cite{Alexandrov:1995kv} - that is, one can reduce the AKSZ theory from the original QP manifold to that of the new manifold. Solutions of this reduced AKSZ theory will look like branes wrapping cycles of the target space. We will describe the relation to AKSZ sigma models in a later section and will describe the wrapping procedure here.

We start from the following ingredients.
\begin{itemize}
    \item An NQP manifold $\Mcal = \Ncal \times \Ycal$ of degree $n\geq 2$ where
    \begin{equation}
        \Ycal = T^{*}[n]T[1]Y
    \end{equation}
    and $\Ncal$ is otherwise generic, with underlying commutative manifold\footnote{By `underlying commutative manifold' we mean the commutative manifold $M$ whose structure sheaf is the sheaf of degree 0 functions on $\Mcal$, i.e. $C^{\infty}(M) = C^{\infty}_{0}(\Mcal)$. This is well defined since we are working on non-commutative manifolds with a non-negative grading. We will also refer to this as the manifold in degree 0.} $N$. The underlying commutative manifold for $\Mcal$ is $M=N\times Y$, a direct product manifold. The symplectic form will be written $\omega_{\Mcal} = \dd\vartheta_{\Mcal}$, where $\vartheta_{\Mcal}$ is the canonical symplectic potential. The induced Poisson Bracket on $\Mcal$ will be written $(\cdot,\cdot)_{\Mcal}$.
    
    \item The Q-structure of $\Mcal$ should be a \textit{lift of the de Rham differential of $Y$}, seen as the vector field $\dr_Y\equiv \xi^m \pd/\pd y^m$ on $T[1]Y$, with respect to the bundle projection $p$ that is the composition $\Ncal\times \Ycal \xrightarrow{\pi_\Ycal} \Ycal\xrightarrow{\pi_{T[1]Y}} T[1]Y$. Explicitly this lift condition means $Q_\Mcal p^\star=p^\star \dr_Y$, which partially determines the form of the hamiltonian $\Theta_\Mcal$ in local coordinates:
    \begin{equation}\label{eq:generic_Q_structure}
        \Theta_{\Mcal} = -\xi^{m} q_{m} + \sum_{k=0}^{n+1}\tfrac{1}{k!}\theta_{m_{1}...m_{k}}(z,y)\xi^{m_{1}}...\xi^{m_{k}}
    \end{equation}
    where $q$ are the degree $n$ conjugate momenta to $y$ on $T^\star[n]T[1]Y$ and $z$ are generic homogeneous coordinates on $\Ncal$. The $\theta_{k} = \theta_{k}(z,y)\xi^{k}$ can be viewed as ($C^{\infty}(\Ncal)$-valued) differential forms on $Y$ and we demand that they must be $\dr_Y$-closed differential forms.
    
    \item A Q manifold $\Xcal = (T[1]X,\dd)$ where $X$ is compact, without boundary, and has dimension $d<n$. $\dd$ is the de Rham differential.
    
    \item A choice of `wrapping map' $\wrap:X\rightarrow Y$, defined up to homotopy.
\end{itemize}

We aim to produce a new NQP manifold $\Wcal$ from $\Mcal,\Xcal$, which describes a brane where $X$ has been wrapped over $Y$ and both cycles have been shrunk. The resulting QP manifold should therefore have degree $n-d$ and underlying commutative manifold $N$. There is a natural choice of manifold of degree $n-d$ given by the mapping space $\Mcal^{\Xcal}:=\mathrm{maps}(\Xcal \rightarrow \Mcal)$. However, this manifold is infinite dimensional. We will see that we can define a coisotropic reduction of $\Mcal^{\Xcal}$ that produces a finite dimensional NQP manifold which only depends on the topology of $X$ and the homotopy class of $\wrap$.

\subsubsection*{Properties of the mapping space}

The infinite dimensional space $\Mcal^{\Xcal}$ consists of maps $f$ which are defined by their pullback action on the coordinates on $\Mcal$. Using generic homogeneous coordinates $Z^{A}$ for $\Mcal$ and coordinates $(\sigma^{\alpha},\dd\sigma^{\alpha})$ for $\Xcal$ adapted to $\dr$ ($\dr (\sigma^\alpha)=\dd\sigma^\alpha,\dd \dr \sigma^\alpha=0$) we have
\begin{equation}\label{eq:functional_coords_as_differential_forms}
    f^{*}Z^{A} = \mathbf{Z}^{A}(\sigma,\dd\sigma) = Z_{0}^{A}(\sigma) + Z_{1\,\alpha}^{A}(\sigma)\dd\sigma^{\alpha}+...+ \tfrac{1}{d!} Z_{d\,\alpha_{1}...\alpha_{d}}^{A}(\sigma)\dd\sigma^{\alpha_{1}} ... \dd\sigma^{\alpha_{d}}
\end{equation}
Defining the components $Z_{k}$ is equivalent to defining the map $f$. To interpret the $Z_k$ we consider a change of coordinates on $\Mcal$ given by $\tilde{Z}^{A} = \tilde{Z}^{A}(Z)$ and note that
\begin{equation}\label{eq:change_of_coordinates}
    \begin{split}
        f^{*}\tilde{Z}^{A}(Z) &= \tilde{Z}^{A}(f^{*}Z) \\
        &= \tilde{Z}^{A}(Z_{0}) + Z_{1\,\alpha}^{B} \dd\sigma^{\alpha} \frac{\partial \tilde{Z}^{A}}{\partial Z^{B}}(Z_{0}) \\
        & \qquad \qquad + \frac{1}{2}\dd\sigma^{\alpha}\dd\sigma^{\beta} \left( Z^{B}_{2\,\alpha\beta}\frac{\partial \tilde{Z}^{A}}{\partial Z^{B}}(Z_{0}) + Z^{B}_{1\, \alpha} Z^{C}_{1\,\beta} \frac{\partial^{2} \tilde{Z}^{A}}{\partial Z^{B} \partial Z^{C}}(Z_{0}) \right) + ...
    \end{split}
\end{equation}
Therefore, in spite of the index structure, these in general are \textbf{not} vector-bundle-valued differential forms, with the exception of $Z_1$ which is an $f_0^\star T\Mcal$-valued 1-form for the map $f_0=f\circ \mathrm{s}_0$, where $\mathrm{s}_0:X\to \Xcal$ is the zero section of $\Xcal=T[1]X$. Of the other components, $Z_0^A$ defines the map $f_0:X\to \Mcal$, while the $Z_{k}^A$ for $k>1$ transform ``affinely'' whenever $Z_{k'}^A\neq 0$ for any $0<k'<k$.\footnote{Exploiting Batchelor's theorem to write $\Mcal$ as a graded vector bundle only improves this situation in that some $Z_0$ take values in a vector bundle as well.} Since we may not set $Z_k^A=0$ consistently in general, this introduces a subtlety for our reduction procedure which we will discuss later in this section.

The QP structure on the mapping space is induced by that on $\Mcal$ through transgression. The symplectic structure is given by
\begin{equation}\label{eq:transgressed_symplectic_form}
    \omega_{\Mcal^{\Xcal}} = \int_{\Xcal}\tfrac{1}{2}\delta \mathbf{Z}^{A}(\omega_{\Mcal})_{AB}\, \delta \mathbf{Z}^{B} = \sum_{k}\int_{\Xcal}\tfrac{1}{2}\delta Z^{A}_{k}(\omega_{\Mcal})_{AB} \delta Z^{B}_{d-k}
\end{equation}
which induces a Poisson bracket $[\cdot,\cdot]$ on $\Mcal^{\Xcal}$. This Poisson bracket can be conveniently expressed in terms of `test functions' as in \cite{Arvanitakis:2021wkt}. Given arbitrary functions $\epsilon, \eta$ on $\Xcal$ --- which correspond to differential forms on $X$ since $\Xcal=T[1]X$ --- they write
\begin{equation}\label{eq:transgressed_PB}
    \left[ \int_{\Xcal} \mathbf{Z}^{A}\epsilon\, ,\, \int_{\Xcal} \mathbf{Z}^{B} \eta \right] =  (-1)^{(B+n)\epsilon+d}\int_{\Xcal} (Z^{A},Z^{B})_{\Mcal}\, \epsilon\eta
\end{equation}
where in the exponent we have used the shorthand $B,\epsilon$ for the degrees of the respective functions. From \eqref{eq:transgressed_symplectic_form} and \eqref{eq:transgressed_PB} we can see that if $Z^{A}$ is dual to $Z^{B}$ on $\Mcal$, then $Z^{A}_{k}$ will be dual to $Z^{B}_{d-k}$ on $\Mcal^{\Xcal}$. Furthermore, if we are working in Darboux coordinates, so that components of $\omega_{\Mcal}$ are constant, then by performing a Hodge decomposition
\be
\label{eq:hodge}
\Omega^{k}(X) = \mathcal{H}^{k} \oplus \dd \Omega^{k-1} \oplus \dd^{\dagger}\Omega^{k+1}
\ee
with respect to some arbitrary metric, exact forms $Z^{A}_{k}$ will be dual to co-exact $Z^{B}_{d-k}$ and harmonic forms will be dual to harmonic forms. For convenience we introduce orthogonal projectors
\be
\label{eq:hodgeprojectors}
P_{\mathcal H},\qquad P_{\ex}\,,\qquad P_{\co}
\ee
onto harmonic, exact, and co-exact forms respectively.

The Q-structure $D$ on $\Mcal^\Xcal$ is defined as the hamiltonian vector field
\be
D=\dr + Q_\Mcal\,,\qquad D = [\Theta_{\Mcal^{\Xcal}},\cdot]\,,
\ee where the hamiltonian is
\begin{equation}
\label{eq:transgressedhamiltonian}
    \begin{split}
        \Theta_{\Mcal^{\Xcal}} &= (-1)^{d}\int_{\Xcal} \bm\Theta_{\Mcal} + (-1)^{d+n+1}\int_{\Xcal}\imath_{\dd} \bm\vartheta_{\Mcal}
    \end{split}
\end{equation}
where each term generates the lift of $Q_{\Mcal}$ and $\dd$ to $\Mcal^{\Xcal}$ respectively. Note that implicit in this formulae is the fact that we have pulled back/transgressed $\Theta_{\Mcal},\vartheta_{\Mcal}$ to objects on $\Xcal$; we have used boldface to highlight this. The signs are such that $D=\dr_\Xcal + Q_\Mcal$.

\subsubsection*{The coisotrope}

We need to perform a coisotropic reduction to obtain a finite dimensional NQP manifold. This is a generalisation of symplectic reduction for Poisson manifolds which requires a coisotropic ideal $\Ical \subset C^{\infty}(\Mcal^{\Xcal})$, i.e. an ideal that satisfies
\begin{equation}
    [\Ical,\Ical] \subseteq \Ical
\end{equation}
The description of the quotient manifold is given in two equivalent ways. In one description, we take the submanifold $\mathcal{C}\subset \Mcal^{\Xcal}$ defined by the vanishing of $\Ical$ and quotient by transformations generated by $\Ical$. Alternatively, we can describe the structure sheaf of the quotient manifold as the normaliser $N(\Ical)$ of $\Ical$, quotiented by $\Ical$. That is
\begin{equation}\label{eq:coiso_reduction}
    \Wcal = \mathcal{C} / [\Ical,\cdot] \qquad \Leftrightarrow \qquad C^{\infty}(\Wcal) = N(\Ical)/\Ical
\end{equation}
Such a manifold has a natural Poisson structure induced from that on the mapping space; see appendix \ref{app:CoisoreductionOfPoisson} for a review. Further, provided the ideal is closed with respect to the Q structure, i.e.~$D\Ical = [\Theta_{\Mcal^{\Xcal}},\Ical]\subseteq \Ical$, the reduced space has a Q-structure induced from the image of the Hamiltonian function under the quotient map:
\begin{equation}
    \Theta_{\Wcal} = \Pi(\Theta_{\Mcal^{\Xcal}}) \qquad \Pi:N(\Ical) \rightarrow N(\Ical)/\Ical
\end{equation}
This closure is precisely the statement that $\Theta_{\Mcal^{\Xcal}} \in N(\Ical)$.

We build our ideal $\Ical = \left< \Ical_{\Ncal},\Ical_{\Ycal} \right>$ in 2 parts, each defining a restriction to some submanifold of $\Mcal^{\Xcal} = \Ncal^{\Xcal}\times \Ycal^{\Xcal}$. This factorisation is convenient because $\Ycal$ may be thought of as `longitudinal' to the cycle to be wrapped, while $\Ncal$ is `transverse'.

 On $\Ycal^{\Xcal}$, we would like the maps in degree 0 to restrict to the fixed wrapping map $\wrap:X\to Y$. This restriction is naturally given by the zero locus of the ideal generated by $\bm{y}-\wrap$ and its closure under $D$. Using \eqref{eq:generic_Q_structure} and \eqref{eq:transgressedhamiltonian}, we find
\begin{equation}
    \Ical_{\Ycal} = \left<\bm{y}-\wrap,\, \bm{\xi} + \dd\wrap \right>
\end{equation}
This is clearly coisotropic in the coordinates on $\Ycal$. The angled brackets $\langle \cdots \rangle$ will always denote the ideal generated by $\cdots$.

On $\Ncal^{\Xcal}$, we follow \cite{Bonechi:2010tbl} and take the coisotropic submanifold to consist --- in the first instance --- of closed maps under the transgressed differential $\bm{\dd}$ on $\Ncal^{\Xcal}$. In degree zero we realise this via a choice of degree preserving embedding $N\hookrightarrow \Ncal^{\Xcal}$. By degree counting this is a map of (ordinary) manifolds $N\hookrightarrow N^X$, and we choose this to be the map sending each $n\in N$ to the constant map $X\to \{n\}$ (which is $\bm{\dr}$-closed). Beyond degree zero, we simply set the coisotropic part of each coordinate in the superfield expansion \eqref{eq:functional_coords_as_differential_forms} to zero (using the Hodge decomposition). Therefore we define $\Ical_\Ncal$ such that\footnote{The reduction by such a coisotropic ideal is related to the reduction by `contractible pairs' in the BV formalism \cite{Barnich:2009jy}.}
\begin{equation}
    \Ical_{\Ncal} \supset \left< P_\mathrm{co}z^{A}_{k} \right>
\end{equation}
for all values of $k$ in the expansion \eqref{eq:functional_coords_as_differential_forms}, where $z^A$ is a generic coordinate on $\Ncal$. 
 If we consider the vanishing locus of $\Ical_\Ncal$ and $\Ical_\Ycal$ simultaneously, we see that we are restricted to $x_{0} = \text{const}$, $y_{0} = \wrap$. This gives an embedding $N \hookrightarrow \Mcal^{\Xcal}$. Similarly, a choice of degree preserving embedding $N\hookrightarrow \Mcal^{\Xcal}$ defines our ideal in degree 0.

This alone is not enough as we would like the reduced manifold $\Wcal$ to be an N-manifold, i.e. a graded manifold with non-negative coordinates, such that in degree zero the structure sheaf is that of an ordinary manifold.\footnote{There are issues not just with negative-graded but also with degree zero `formal' coordinates; see e.g.~\cite[section 2]{Arvanitakis:2022npu}.} To remove these, we include harmonic generators of the maps $z^{A}_{k}$ for maps such that $\deg z^{A} - k \leq 0$. The exception to this is the maps $x_{0}$ for which we do not include the harmonic (i.e. constant map) representatives. We therefore have
\begin{equation}
\label{eq:Ical_Ncal}
    \Ical_{\Ncal} = \left<\begin{cases} P_\mathrm{co}z^{A}_{k} ,\, P_\mathcal{H}z^{A}_{k'}\,|\, \forall k'\geq \deg z^{A} \quad 
    &\text{if}\quad \deg z^A>0 \\ P_\mathrm{co}z^{A}_{k}\,, P_\mathcal{H}z^{A}_{k'}\,|\, \forall k'>0\quad &\text{if}\quad \deg z^A=0 \end{cases}  \right>
\end{equation}

To see that this is coisotropic, we use \eqref{eq:transgressed_PB} and the surrounding discussion to note that the co-exact generators are dual to exact generators. Hence, these terms are coisotropic with respect to all of $\Ical_{\Ncal}$. The harmonic generators could be dual to some other harmonic generator in $\Ical_{\Ncal}$. However, since we only include harmonic $z^{A}_{k}$ for $0\geq \deg z^{A} - k$, the dual coordinate $z^{B}_{k'}$ on $\Mcal^{\Xcal}$ has
\begin{equation}
    \begin{split}
        \deg z^{B} - k' &= n- \deg z^{A} - (d-k) \\
        &= (n-d) - (\deg z^{A} - k) \\
        &> 0
    \end{split}
\end{equation}
so is not included in $\Ical_{\Ncal}$. The total ideal $\Ical = \left<\Ical_{\Ncal},\,\Ical_{\Ycal}\right>$ is coisotropic, as required.

\subsubsection*{Metric and coordinate independence}

The construction of the ideal $\Ical$ appears to rely on a choice of metric on $X$ but we claim that the resulting reduced manifold depends on only the topological data of $X$. This can be best seen from the vanishing locus $\Ccal\subset \Mcal^{\Xcal}$. This is given by maps which are either $\dr$-closed (for $\deg z^A=0$ and some values of $k,k'$) or ones which are also $\dr$-exact (in all other cases), as specified in \eqref{eq:Ical_Ncal}. The metric only appears in the specification of the vanishing ideal that represents $\Ccal$ but $\Ccal$ does not in itself depend on metric data. (This apparent metric dependence thus may perhaps be seen as due to `gauge-fixing'.)

Our construction also appears to depend on a choice of coordinates on $\Ncal$. To see that this is well defined, we will show that the submanifold $\Ccal$ is invariant under a change of coordinates $\tilde{z}^{A} = \tilde{z}^{A}(z)$. Using formula \eqref{eq:change_of_coordinates}, we can write the $k$-form component of the transgressed $\tilde{\bm{z}}$ as
\begin{equation}
    \tilde{z}^{A}_{k} \sim \sum_{j} C^{A}{}_{A_{1}...A_{j}}(z_{0}) z^{A_{1}}_{k_{1}}...z^{A_{j}}_{k_{j}}
\end{equation}
such that $k_{1}+...+k_{j} = k$ and $\deg z^{A_{1}}+...+\deg z^{A_{j}}\leq \deg \tilde z^A_k$; here we emphasise that the last inequality holds true because $\Mcal$ --- and thus $\Ncal$ --- \emph{was assumed to be an N-manifold} (its structure sheaf is non-negatively graded).  Restricting to $\Ccal$, the coordinates $z^{A_{i}}_{k_{i}}$ are all closed under $\dd$ and hence so is $\tilde{z}^{A}_{k}$. Further, if $\deg\tilde{z}^{A} - k \leq 0$, then at least one of $\deg z^{A_{i}} - k_{i}\leq 0$. This means that $z^{A_{i}}_{k_{i}}$ is exact. A product of closed and exact forms is exact and hence so is $\tilde{z}^{A}_{k}$ as required.

\subsubsection*{Closure under $D$}

The final condition to check is that the ideal $\Ical$ is closed under the Q-structure $D = [\Theta_{\Mcal^{\Xcal}},\cdot]$ on $\Mcal^{\Xcal}$. We have already checked that $\Ical_{\Ycal}$ is $D$-closed and so we need only check how $D$ acts on the generating coordinates of $\Ical_{\Ncal}$. We can use formula \eqref{eq:transgressed_PB} with  $P_\mathrm{co}\epsilon=0$. This choice of epsilon selects out the harmonic and co-exact generators $z^{A}_{k}$, respectively. We have
\begin{equation}\label{eq:closure_of_ideal}
    \begin{split}
        \left[\Theta_{\Mcal^{\Xcal}},\int_{\Xcal}\bm z^{A} \epsilon\right]&= (-1)^{d}\left[\int_{\Xcal} \bm{\Theta}_{\Mcal}, \int_{\Xcal}\bm{z}^{A}\epsilon \right] + (-1)^{d+n+1}\left[ \int_{\Xcal}\imath_{\dd} \bm{\vartheta}_{\Mcal}, \bm{z}^{A}\epsilon \right] \\
        &= \int_{\Xcal} \bm{(\Theta_{\Mcal},z^{A})_{\Mcal}}\, \epsilon + \int_{\Xcal}\dd\bm{z}^{A}\epsilon
    \end{split}
\end{equation}

The second term vanishes when $\epsilon$ is closed. We therefore need to only consider the first term. We can see whether this term is contained within $\Ical$ by transgressing the function $(\Theta_{\Mcal},z^{A})_{\Mcal}$ to the mapping space and evaluating it over $\Ccal$. If the integral vanishes when integrated against all closed $\epsilon$ then the ideal is closed under $D$. Using the form of the Hamiltonian function we find
\begin{equation}
    (\Theta_{\Mcal},z^{A})_{\Mcal} \propto (\omega_{\Mcal})^{AB}\sum_{k} \frac{\partial \theta_{k}(z,y)}{\partial z^{B}} \xi^{k}
\end{equation}
Transgressing this to the mapping space and restricting to $\Ccal$, we replace $y\to \wrap$, $\xi\to \dd\wrap$, $z\to \bm{z}$ for $\bm{z}$ closed (or exact). Integrating this against $\epsilon$ we get
\begin{equation}
    \left. \left[ \Theta_{\Mcal^{\Xcal}}, \int_{\Xcal} \bm{z}^{A}\epsilon \right] \right|_{\Ccal} \propto \int_{\Xcal} \left( (\omega_{\Mcal})^{AB} \wrap^{*} \sum_{k} \frac{\partial \theta_{k}}{\partial z^{B}}(\bm{z}) \right)\epsilon
\end{equation}

We require this to vanish for the above $\epsilon$. When $\epsilon$ is exact, this indeed vanishes if we impose $\dr_Y \theta_k=0$.  If $\epsilon$ is harmonic, however, we find constraints on the coefficients $\theta_{k}$ that we address case-by-case, in general.

\subsubsection*{The reduction, metric independence, and homotopy invariance}

Given the coisotropic reduction $\Ical$, we consider the reduction given in \eqref{eq:coiso_reduction}. We will consider the structure sheaf construction of the reduced manifold. The normaliser $\Ncal(\Ical)$ of the ideal is generated by 
\begin{equation}
    N(\Ical) \sim \left \{ P_{\mathcal H} x_{0} ,\, P_{\mathcal H} z^{A}_{k},\, \Ical\,|\, 0< \deg z^{A} - k \leq n-d  \right\}
\end{equation}
We can expand the $P_{\mathcal{H}}z^{A}_{k} = z^{A,a}_{k} e_{a}$ in some basis $\{e_{a}\}$ of $\mathcal{H}^{k}$, so the $z^{A,a}_{k}$ are constant parameters of degree $\deg z^{A} - k$. In the case that the respective cohomology group is 1 dimensional (e.g. for $\mathcal{H}^{0}, \mathcal{H}^{d}$) we will omit the $a$ index and simply identify e.g. $z^{A}_{d} = z^{A}_{d}\mathrm{vol}_{X}$. We see that the structure sheaf $C^{\infty}(\Wcal) = N(\Ical)/\Ical$ is given therefore generated by
\begin{equation}
    C^{\infty}(\Wcal) = N(\Ical)/\Ical \sim \{x_{0},\, z^{A,a}_{k}\,|\, 0< \deg z^{A} - k \leq n-d \}
\end{equation}
That is, the structure sheaf is given by all smooth functions in the $z^{A,a}_{k}$ (and $x_{0}$).

The Hamiltonian function on $\Wcal$ is given by the projection $\Pi:N(\Ical)\rightarrow N(\Ical)/\Ical$ of $\Theta_{\Mcal^{\Xcal}}$. We will confirm in the examples that the final result is given by
\begin{equation}\label{eq:reduced_Theta}
    \begin{split}
        \Theta_{\Wcal} &= \Pi(\Theta_{\Mcal^{\Xcal}}) = \int_{\Xcal} \sum_{k}(-1)^{k}\wrap^{*}\theta_{k}(z)
    \end{split}
\end{equation}
where the $z$ are now the harmonic representatives of the cohomology groups on $X$. Expanding the harmonic $z^{A}_{k}$ in terms of the constant coordinates $z^{A,a}_{k}$, we can perform the integral over $X$ with the convention that the volume form is on the right of the integrand, so we pull out constants from the left. Once this is done the final result will no longer be an integral but will be a function in the $z^{A,a}_{k}$ which will involve, in general, a sum over cohomology groups, which will be discrete in all cases (we only consider wrapping over compact cycles).

Formula \eqref{eq:reduced_Theta} seems to depend on some metric to choose the harmonic representatives for the $z$. However, under a change of metric, the harmonic representatives change by a $\dd$-exact term, and since we have assumed that the forms $\theta_{k}$ are closed, this shift will not change the integral. Furthermore, since the forms $\theta_{k}$ are closed, the evaluation of the integral only depends on the homotopy class of $\wrap:X\to Y$. Therefore the construction is metric-independent and homotopy invariant.

Now that we have defined the reduction in complete generality, we will see many examples of how this works in practice. There are three interesting cases to consider.
\begin{enumerate}
    \item $\dim X = 0$ -- The process effectively shrinks  $Y$ to a point.
    
    \item $\dim Y = 0$ -- We produce a QP manifold with the same underlying commutative manifold but with a different degree. 
    
    \item $X = Y$ -- We produce a QP manifold which corresponds to a brane wrapping the internal manifold.
\end{enumerate}

\section{Example -- $\dim X = 0$}\label{sec:dimX=0}

We consider first a simple example to show that in the simple case that $\dim X = 0$, our procedure effectively reduces to dimensional reduction on $Y$. Consider the ingredients
\begin{equation}
    \Mcal = T^{*}[n]T[1] (N\times Y) \qquad X = \mathrm{pt}
\end{equation}
Taking $\Ncal = T^{*}[n]T[1]N$, $\Ycal = T^{*}[n]T[1]Y$ and $\Xcal = T[1]X = \mathrm{pt}$, we introduce the Darboux coordinates
\begin{equation}
    \arraycolsep = 5pt
    \begin{array}{ccc}
        \Ncal & \qquad & \Ycal \\
        \begin{array}{r|cccc}
            \text{coord} & x^{\mu} & \psi^{\mu} & \chi_{\mu} & p_{\mu}  \\ \hline
            \text{deg} & 0 & 1 & n-1 & n
        \end{array} & & \begin{array}{r|cccc}
            \text{coord} & y^{m} & \xi^{m} & \phi_{m} & q_{m}  \\ \hline
            \text{deg} & 0 & 1 & n-1 & n
        \end{array}
    \end{array}
\end{equation}
We will take the QP structure to be given by the symplectic form and Hamiltonian function
\begin{align}
    \omega_{\Mcal} &= \dd p \,\dd x + \dd q\, \dd y - \dd \psi\, \dd\chi - \dd\xi\, \dd\phi \label{eq:n_brane_symplectic_form} \\
    \Theta_{\Mcal} &= -\psi p - \xi q + \tfrac{1}{n!}F_{n}\psi^{n} + \tfrac{1}{(n-1)!} F_{n-1}\psi^{n-1}\xi + ... + \tfrac{1}{d!(n-d)!} F_{n-d}\psi^{n-d}\xi^{d}
\end{align}
We have suppressed all indices but they should be read as being contracted in the natural way. The coefficients $F_{k}$ can be thought of as elements of $\Omega^{k}(N)\times \Omega^{n-k}(Y)$. These should be closed under the differential $\dd_{Y}$ on $Y$. So for example, $F_{n}\psi^{n} = F_{n}(x,y)_{\mu_{1}...\mu_{n}}\psi^{\mu_{1}}...\psi^{\mu_{n}}$ should be viewed as a differential $n$-form on $N$, but a constant function on $Y$. In the ansatz above, we have assumed a trivial connection on the bundle. We can easily reintroduce it by making the replacement $\xi \rightarrow \mathcal{A} = \xi + A\psi$, where $A$ is the connection, however it won't change our final result so we omit it for simplicity.

The first step in the reduction process is to transgress the QP structure to $\Mcal^{\Xcal}$. But since $\Xcal$ is 0 dimensional, we have $\Mcal^{\Xcal} \simeq \Mcal$. Next, we need to choose a wrapping map $\wrap:X=\text{pt}\to Y$, or equivalently, a (degree preserving) embedding $N \hookrightarrow \Mcal^{\Xcal} \simeq \Mcal$. This is equivalent to choosing some point $\hat{y} \in Y$ and defining the embedding $N\to (N,\hat{y}) \subset M$. This is described by the ideal
\begin{equation}
    \Ical_{0} = \left<y^{m}-\hat{y}^{m}\right>
\end{equation}
We then want to form the closure of this ideal with respect to differential $Q$ on $\Mcal$. We get
\begin{equation}
    \begin{split}
        \Ical_{\Ycal} = \left< \Ical_{0}, Q\Ical_{0} \right>  =\left< y^{m}-\hat{y}^{m}, \, \xi^{m} \right>
    \end{split}
\end{equation}
It is easy to check from \eqref{eq:n_brane_symplectic_form} that this is indeed coisotropic with respect to the Poisson bracket on $\Mcal$. In principal, we also need to restrict the maps into $\Ncal$ to those that are closed/exact with respect to $\dd$ on $X$. However, since $\dim X = 0$, this is a trivial constraint and so we just have $\Ical = \Ical_{\Ycal}$.

To perform the coisotropic reduction, we need to go to first find the normaliser $N(\Ical)$ of $\Ical$, which can easily be verified to be generated by the coordinates
\begin{equation}
    N(\Ical) \sim \{ x^{\mu},\psi^{\mu},\chi_{\mu},p_{\mu},y^{m}-\hat{y}^{m}, \xi^{m} \}
\end{equation}
The structure sheaf of the new QP manifold $\Wcal$ is then defined to be the quotient of this by the ideal $\Ical$. That is $C^{\infty}(\Wcal) = N(\Ical)/\Ical$, which is generated by
\begin{equation}
    N(\Ical)/\Ical \sim \{ x^{\mu},\psi^{\mu},\chi_{\mu},p_{\mu} \} \quad \Rightarrow \quad \Wcal = T^{*}[n]T[1]N
\end{equation}
Note that, by construction $\Theta_{\Mcal} \in N(\Ical)$ and so we can find the new Hamiltonian function through the natural projection $\Pi:N(\Ical) \rightarrow N(\Ical)/\Ical$, which gives
\begin{equation}
    \Theta_{\Wcal} = \Pi(\Theta_{\Mcal}) = -\psi p + \tfrac{1}{n!}F_{n}(x,\hat{y})\psi^{n}
\end{equation}
and the final symplectic form is
\begin{equation}
    \omega_{\Wcal} = \dd p\, \dd x - \dd\psi\, \dd\chi
\end{equation}

We see that this procedure has produced a new QP manifold with the same degree but with underlying commutative manifold $N$. We see that we have effectively collapsed $Y$ to the point $\hat y$. In the case where $Y$ is a Lie group, we find the same result as in symplectic reduction modulo $T[1]Y$ \cite{Arvanitakis:2021lwo}. If we were to choose a different wrapping map $\wrap':X\mapsto \hat{y}'$ that is homotopic to $\wrap:X\mapsto \hat{y}$, then we end up with the same graded manifold where the Q-structure is evaluated for $F_{n}(x,\hat{y}')$. However, the condition that the $F_{n}$ is closed on $Y$ says that it is constant, and hence the Q-structures are the same. This demonstrates the homotopy invariance of our construction.

\section{Examples -- $\dim Y = 0$}\label{sec:dimY=0}

\subsection{$n$-brane $\to$ $(n-1)$-brane}\label{sec:n-brane}

Let's now consider the same example as above, but instead of having $\dim X = 0$, we will take the dimension of the fibre $\dim Y = 0$ and take $X$ to be non-trivial. We will take the ingredients
\begin{equation}
    \Mcal = T^{*}[n]T[1]M \qquad X = S^{1}
\end{equation}
We will use the homogeneous coordinates coordinates
\begin{equation}
\arraycolsep = 5pt
    \begin{array}{c}
    \Mcal \\
    \begin{array}{r|cccc}
        \text{coord} & x^{\mu} & \psi^{\mu} & \chi_{\mu} & p_{\mu}  \\ \hline
         \text{deg} & 0 & 1 & n-1 & n
    \end{array}
    \end{array}
\end{equation}
and use the coordinates $\sigma,\dd\sigma$ on $\Xcal = T[1]S^1$. The Hamiltonian function and symplectic form are given by
\begin{align}
    \omega_{\Mcal} &= \dd p\,\dd x - \dd\psi\, \dd\chi \\
    \Theta_{\Mcal} &= -\psi p + \tfrac{1}{n!}F_{n} \psi^{n}
\end{align}
Since $Y=\text{pt}$ in this example, we do not need to impose any constraints on the coefficients $F_{n}$.

We need to transgress this structure to the mapping space $\Mcal^{\Xcal}$. This is now an infinite dimensional graded manifold whose points $f\in \Mcal^{\Xcal}$ can be described by their pullback action on coordinates on $\Mcal$. That is, we have
\begin{equation}\label{eq:n-brane_S1_coordinate_pullback}
    f^{*} Z^{A} = \bm{Z}^{A}(\sigma,\dd\sigma) = Z^{A}_{0}(\sigma) + Z^{A}_{1}(\sigma)\dd\sigma\,.
\end{equation}
The transgressed Hamiltonian function is given by
\begin{equation}\label{eq:n-brane_S1_Q-structure}
    \begin{split}
        \Theta_{\Mcal^{\Xcal}} &= (-1)^{1}\int_{\Xcal}\bm{\Theta}_{\Mcal} + (-1)^{n-1+1}\int_{\Xcal}\imath_{\dd} \bm{\vartheta}_{\Mcal} \\
        &= -\int_{T[1]S^{1}}-\bm{\psi p}+\tfrac{1}{n!}F_{n}(\bm{x})\bm{\psi^{n}} + (-1)^{n}\int_{T[1]S^{1}} \bm{p}\dd \bm{x} - \tfrac{1}{n}(\bm{\psi} \dd\bm{\chi} + (n-1) \bm{\chi}\dd\bm{\psi})
    \end{split}
\end{equation}
The bold-faced letters in the expression correspond to functions pulled back to functions on $\Xcal$ as in \eqref{eq:n-brane_S1_coordinate_pullback}. The Berezin integral over $T[1]S^{1}$ selects the maximal degree component of the integrand (i.e. the 1-form components) and integrates it over $S^{1}$. Our convention is that we normalise with an overall factor of $\mathrm{vol}(S^{1})$, and so for the flat metric on $S^{1}$ we have
\begin{equation}
    \int_{T[1]S^{1}} ... = \frac{1}{2\pi}\int_{S^{1}} (...)_{1}
\end{equation}

The next step is to define the coisotropic ideal $\Ical = \left<\Ical_{\Ncal}, \Ical_{\Ycal}\right>$. Since $Y$ is trivial, so is the ideal $\Ical_{\Ycal}$ and hence we need only determine $\Ical_{\Ncal}$. Following section \ref{sec:wrapping_QP}, we first start by restricting to all closed maps. That is, we take
\begin{equation}
    \Ical_{\Ncal} \supset \left< P_\mathrm{co}x_{k} ,\, P_\mathrm{co}\psi_{k},\, P_\mathrm{co}\chi_{k},\, P_\mathrm{co}p_{k} \right>
\end{equation}
To define this ideal we choose some arbitrary metric on $S^{1}$, and for simplicity we can take the flat metric. We then also add the harmonic representatives for $Z^{A}_{k}$ such that $\deg Z^{A} - k \leq 0$ (except for $x_{0}$). This gives
\begin{equation}
    \Ical_{\Ncal} = \left< P_\mathrm{co}x_{0},\, P_\mathrm{co}\psi_{0},\, P_\mathrm{co}\chi_{0},\,P_\mathrm{co} p_{0} ,\, P_\mathcal{H}x_{1}, P_\mathcal{H}\psi_{1} \right>
\end{equation}
Again, in degree 0, the vanishing locus of this ideal restricts us to maps $x_{0} = \text{const}$ and hence defines a natural embedding $M\hookrightarrow \Mcal^{\Xcal}$. It is a quick check using \eqref{eq:transgressed_PB} that this ideal is coisotropic. Indeed, the Poisson bracket of the co-exact generators with any other generators will vanish, as they are dual to exact maps. The harmonic $x_{1},\psi_{1}$ representatives are dual to $p_{0},\chi_{0} \in \mathcal{H}^{0}$ respectively and these do not appear in the generating set of $\Ical_{\Ncal}$.

We will also verify that this ideal is closed with respect to the Q-structure \eqref{eq:n-brane_S1_Q-structure}. Using the test function form of the Poisson bracket \eqref{eq:transgressed_PB}, we can calculate $D$ acting on the generators by calculating
\begin{equation}
    \left[ \Theta_{\Mcal^{\Xcal}}, \int_{T[1]S^{1}} \bm{Z}^{A}\epsilon \right] = -\left[ \int_{T[1]S^{1}} \bm{\Theta}_{\Mcal}, \int_{T[1]S^{1}} \bm{Z}^{A}\epsilon \right] + (-1)^{n} \left[ \int_{T[1]S^{1}} \imath_{\dd}\bm{\vartheta}_{\Mcal}, \int_{T[1]S^{1}} \bm{Z}^{A} \epsilon  \right]
\end{equation}
where $\epsilon$ is a function on $\Ncal$ that is closed under $\dd$. Taking $\epsilon \in \mathcal{H}^{k}$ selects the harmonic representative $Z_{1-k}\in \mathcal{H}^{1-k}$, while taking $\epsilon$ to be exact selects the co-exact representative of $Z^{A}_{0}$. The second term gives us
\begin{equation}\label{eq:n-brane_d_exact_PB_term}
    \left[ \int_{T[1]S^{1}} \imath_{\dd}\bm{\vartheta}_{\Mcal}, \int_{T[1]S^{1}} \bm{Z}^{A} \epsilon  \right] \propto \int_{T[1]S^{1}} \dd \bm{Z} \,\epsilon = \frac{1}{2\pi} \int_{S^{1}}\dd Z^{A}_{0} \, \epsilon_{0}
\end{equation}
Taking $\epsilon$ to be closed tells us that $\epsilon_{0}$ is constant. The integrand on the right hand side is therefore exact and so the integral vanishes. The Poisson bracket is then determined by the first term alone which is proportional to
\begin{equation}\label{eq:n-brane_S1_coiso_PB}
    \left[ \Theta_{\Mcal^{\Xcal}}, \int_{T[1]S^{1}} \bm{Z}^{A}\epsilon \right]  \propto \int_{T[1]S^{1}} \bm{(\Theta_{\Mcal}, Z^{A})_{\Mcal}} \, \epsilon
\end{equation}
where the function $(\Theta_{\Mcal},Z^{A})$ is transgressed to the mapping space. We can use these results to confirm that $\Theta_{\Mcal^\Xcal}$ lies always in $\Ical_{\Ncal}$ as outlined in Section \ref{sec:wrapping_QP}. The only non-trivial checks are for the harmonic generators $x_{1}, \psi_{1}$, for which we take $\epsilon=\epsilon_{0}$ to be constant. We have
\begin{equation}
    (\Theta_{\Mcal}, x)_{\Mcal} = \psi\,, \qquad \qquad (\Theta_{\Mcal},\psi)_{\Mcal} = 0
\end{equation}
Transgressing these functions and evaluating on $\Ccal$, we take $\psi_{1}$ to be exact. Hence, both vanish under the integral \eqref{eq:n-brane_S1_coiso_PB} when $\epsilon=\epsilon_{0}$ is constant. This proves that the ideal is closed under $D$.

Now that we have our coisotropic ideal, we perform the coisotropic reduction. The normaliser of $\Ical$ is generated by all the coordinates that are not dual to those in $\Ical$. 
\begin{equation}
    N(\Ical) \sim \{ P_{\Hcal}x_{0},\, P_{\Hcal}\psi_{0},\, P_{\Hcal} \chi_{1},\,P_{\Hcal} p_{1},\, \Ical \}
\end{equation}
The structure sheaf for $\Wcal$ is then $N(\Ical)/\Ical$ which is generated by
\begin{equation}
    N(\Ical)/\Ical \sim \{x_{0},\,\psi_{0},\, \chi_{1},\, p_{1} \} \quad \Rightarrow \quad \Wcal = T^{*}[n-1]T[1]M
\end{equation}
(Note in the expression above we are now working in the coordinates $z_k^{A,a}$ described in section \ref{sec:wrapping_QP}:  $P_{\Hcal} x_{0} = x_{0} \cdot 1$ and $P_{\Hcal} p_{1} = p_{1} \mathrm{vol}$.)
Thus we restrict to harmonic functions for $x,\psi$ --- so they retain their original degrees --- while we restrict to harmonic 1-forms for $\chi, p$ hence they have their degrees shifted down by 1. We therefore end up with the manifold $T^{*}[n-1]T[1]M$.

To find the symplectic form, we use the Poisson brackets \eqref{eq:transgressed_PB} with the $\epsilon, \eta$ appropriate harmonic representatives. We find that
\begin{equation}
    \omega_{\Wcal} = -\dd p_{1}\, \dd x_{0} - \dd\psi_{0} \,\dd\chi_{1}
\end{equation}
To find the form of the Hamiltonian function we project $\Theta_{\Mcal^{\Xcal}}$ under $\Pi:N(\Ical) \rightarrow N(\Ical)/\Ical$. By restricting all coordinates to the harmonic representatives on which $\dd = 0$, we find $\Pi(\imath_{\bar{\dd}}\bar{\vartheta}) = 0$. The term $\tfrac{1}{n!}F_{n}\psi^{n}$ gets projected to $\tfrac{1}{n!}F_{n}(x_{0})\psi_{0}^{n}$ which is a function on $S^{1}$ and hence vanishes under the Berezin integral. We find that we are left with\footnote{Our conventions are that we integrate with the volume form on the right of the integrand, and so we pull constants out from the left. This gives the overall sign.}
\begin{equation}
    \Theta_{\Wcal} = \Pi(\Theta_{\Mcal^{\Xcal}}) = \psi_{0}p_{1}
\end{equation}
Making the change of coordinates $p_{1} \to -p_{1}$ puts the QP manifold in the canonical form for a $(n-1)$-brane. Interestingly, all flux twisting drops out of the Hamiltonian function in this case. This is what happens in the zero-wrapping sector of wrapped branes where physically one ends up with a tensionless brane \cite{Townsend:1996xj}. These are somewhat pathological and hence the physical interpretation of such reductions is less clear. We will see that one can get more interesting reductions if one allows $X$ to wrap some part of $M$.

\subsection{From Courant to Poisson}\label{sec:Courant->Poisson}

Using the formulation set out, we can already find interesting relations between different QP manifolds. Suppose $M$ is a Poisson manifold with Poisson bivector $\pi$. There are two distinct ways to realise this structure as a QP structure. Firstly, we can take the straight cotangent lift of $\pi$ to obtain the following QP manifold 
\begin{equation}
\Wcal = T^{*}[1]M \qquad 
\arraycolsep = 5pt
    \begin{array}{r|cc}
    \text{coord} & \tilde{x} & \tilde{p} \\ \hline
    \text{deg} & 0 & 1
    \end{array}
\qquad
\arraycolsep = 2pt
\begin{array}{rcl}
\omega_{\Wcal} &=& \dd\tilde{p}\,\dd\tilde{x} \\
\Theta_{\Wcal} &=& \tfrac{1}{2}\pi \tilde{p}^{2}
\end{array}
\end{equation}
A quick calculation shows that $(\Theta_{\Wcal},\Theta_{\Wcal}) = 0$ if and only if $\pi$ is Poisson.

Alternatively, we can consider the Lie algebroid structure on $T^{*}M$ whose anchor map is given by the bivector $\pi:T^{*}M\rightarrow TM$ and whose bracket is given by
\begin{equation}
    [\alpha,\beta] = \mathcal{L}_{\pi(\alpha)}\beta - \imath_{\pi(\beta)} \dd\alpha
\end{equation}
This Lie algebroid can be lifted to a Dirac structure $L=(1+\pi)T^{*}M$ within the Courant algebroid $TM\oplus T^{*}M$. Such a structure can be described by a QP manifold via
\begin{equation}
\arraycolsep = 5pt
    \Mcal = T^{*}[2]T[1]M \qquad
    \begin{array}{r|cccc}
    \text{coord} & x & \psi & \chi & p \\ \hline
    \text{deg} & 0 & 1 & 1 & 2
    \end{array} \qquad
    \arraycolsep = 2pt
    \begin{array}{rcl}
        \omega_{\Mcal} &=& \dd p\,\dd x - \dd \psi\, \dd\chi \\
         \Theta_{\Mcal} &=& -\pi p \chi + \tfrac{1}{2}\partial \pi \psi\chi^{2}
   \end{array} 
\end{equation}
Once again $(\Theta_{\Mcal},\Theta_{\Mcal}) = 0$ if and only if $\pi$ is Poisson. We have suppressed indices for convenience. We want to see how, if at all, these constructions are related.

Let us perform a circle reduction of $\Mcal$ as above. We transgress the structure to $\Mcal^{\Xcal}$ where $\Xcal = T[1]S^{1}$. As before, we define $\Ical = \Ical_{\Ncal}$ by first including all co-exact generators
\begin{equation}
    \Ical \supset \left< P_{\co}x_{0},\, P_{\co}\psi_{0},\, P_{\co}\chi_{0},\, P_{\co}p_{0} \right>
\end{equation}
Then we include harmonic representatives to remove coordinates of 0 or negative degree. We will slightly relax the construction set out in section \ref{sec:wrapping_QP} by allowing some new coordinates of degree 0.\footnote{The construction, as set out previously, would still work in this case but we would end up with a trivial Q-structure. To result in a QP manifold with non-trivial Q-structure, we will need to perform an intermediate step before removing the additional degree 0 coordinates.} We will define
\begin{equation}
    \Ical = \left< P_{\co}x_{0},\, P_{\co}\psi_{0},\, P_{\co}\chi_{0},\, P_{\co}p_{0},\, P_{\Hcal}x_{1},\, P_{\Hcal}\chi_{1} \right>
\end{equation}
As before, this ideal is coisotropic.

We will check the closure of this ideal with respect to the Q-structure $D$ on $\Mcal^{\Xcal}$. The transgressed Hamiltonian function is
\begin{equation}
    \begin{split}
        \Theta_{\Mcal^{\Xcal}} &= - \int_{T[1]S^{1}}\bm{\Theta}_{\Mcal} + \int_{T[1]S^{1}}\imath_{\dd}\bm{\vartheta}_{\Mcal} \\
        &= -\int_{T[1]S^{1}}-\pi(\bm{x}) \bm{p\chi} + \tfrac{1}{2}\partial \pi(\bm{x}) \bm{\psi\chi}^{2} + \int_{T[1]S^{1}}\bm{p}\,\dd \bm{x} - \tfrac{1}{2}(\bm{\psi}\dd\bm{\chi} + \bm{\chi}\dd\bm{\psi})
    \end{split}
\end{equation}
We then act with this on $\int \mathbf{Z}^{A} \epsilon$ for some test function $\epsilon$ that must be harmonic, or exact. As in \eqref{eq:n-brane_S1_coiso_PB}, the only non-trivial constraint to check is for the harmonic representatives. We need to check if the following vanishes
\begin{equation}
    \int_{T[1]S^{1}}\bm{(\Theta_{\Mcal},Z^{A})_{\Mcal}}\, \epsilon
\end{equation}
whenever the function $(\Theta_{\Mcal},Z^{A})$ is transgressed and evaluated on $\Ccal$, and if $\epsilon$ is harmonic. Since the only harmonic generators of $\Ical$ are $x_{1},\chi_{1}$, we calculate
\begin{equation}
    (\Theta_{\Mcal},x) = \pi(x)\chi\, , \qquad \qquad (\Theta_{\Mcal},\chi_{\mu}) = \tfrac{1}{2}\partial\pi(x)\chi^{2}
\end{equation}
We transgress these functions to the mapping space and evaluate on the vanishing locus $\Ccal$ of $\Ical$. Noting that these are functions of $x,\chi$ alone, evaluating them on $\Ccal$ means that the zero-form component must be constant functions on $X$, while the 1-form component must be an exact form. Integrating these against a constant function $\epsilon = \epsilon_{0}$ selects the 1-form component, which is exact and hence the integral vanishes as required.

The next step is to perform the coisotropic reduction with respect to this ideal. The normaliser is generated by all coordinates not dual to those in $\Ical$.
\begin{equation}
    N(\Ical) \sim \{ P_{\Hcal}x_{0},\, P_{\Hcal}\chi_{0} ,\, P_{\Hcal}\psi_{1},\, P_{\Hcal}p_{1},\, \Ical \}
\end{equation}
and so we obtain the structure sheaf $C^{\infty}(\widetilde{\Mcal}) = N(\Ical)/\Ical$ which is generated by
\begin{equation}
    N(\Ical)/\Ical \sim \{ x_{0},\,\psi_{1},\,\chi_{0},\, p_{1}\} \quad \Rightarrow \quad \widetilde{\Mcal} = T^{*}[1]TM
\end{equation}
This time, we restrict to harmonic functions for $x,\chi$ so they retain their degree, while we take harmonic 1-forms for $\psi,p$ and hence their degree is shifted down by 1. The resulting Hamiltonian function is $\Pi(\Theta_{\Mcal^{\Xcal}})$ and the symplectic form is derived from the Poisson brackets \eqref{eq:transgressed_PB} with harmonic representatives for $\epsilon,\eta$.
\begin{align}
    \Theta_{\widetilde{\Mcal}} & = -\pi p_{1}\chi_{0} -\tfrac{1}{2}\partial\pi \psi_{1}\chi_{0}^{2} \\
    \omega_{\widetilde{\Mcal}} &= -\dd p_{1} \dd x_{0} + \dd\psi_{1} \dd\chi_{0}
\end{align}
We performed the change of coordinates $p_{1}\rightarrow -p_{1}$, $\chi_{0} \rightarrow -\chi_{0}$ to remove minus signs.

We have arrived at a `halfway house' QP manifold $\widetilde{\Mcal}$. Interestingly, this is the cotangent lift of the complete lift of the Poisson structure $\pi$ on $M$ to the tangent bundle $(TM,\pi^{c})$ \cite{mitric2003poisson}. That is, given any Poisson structure $(M,\pi)$ we define a Poisson structure $(TM,\pi^{c})$ by
\begin{equation}
    \pi^{c} = \pi^{\mu\nu}\frac{\partial}{\partial x_{0}^{\mu}} \frac{\partial}{\partial \psi_{1}^{\nu}} + \tfrac{1}{2}\psi_{1}^{\rho}\partial_{\rho} \pi^{\mu\nu}\frac{\partial}{\partial \psi_{1}^{\mu}} \frac{\partial}{\partial \psi_{1}^{\nu}}
\end{equation}
where $x_{0}$ are coordinates on $M$ and $\psi_{1}$ are coordinates along the vector bundle fibres. We can reduce the QP manifold $\widetilde{\Mcal}$ further by following \cite{mitric2003poisson}. Given any (torsionless) connection on $M$, we can define a global vector field on $TM$ given by the geodesic spray
\begin{equation}
    s = \psi_{1}^{\mu}\frac{\partial}{\partial x_{0}^{\mu}} - \psi^{\mu}_{1} \psi_{1}^{\nu}\Gamma_{\mu\nu}^{\rho}\frac{\partial}{\partial \psi_{1}^{\rho}}\,.
\end{equation}
This has a cotangent lift to $T^{*}[1]TM$ whose hamiltonian is
\begin{equation}
    S = \psi_{1}p_{1} - \Gamma \psi_{1}^{2}\chi_{0}
\end{equation}
From this, we define a new Hamiltonian function
\begin{equation}
\begin{split}
    \Theta'_{\widetilde{\Mcal}} &= -\tfrac{1}{2}(S,\Theta_{\widetilde{\Mcal}})= \tfrac{1}{2}\pi p_{1}^{2} + \psi_{1} f(x_{0},\psi_{1},\chi_{0},p_{1})
\end{split}
\end{equation}
where $f$ is some function of the coordinates whose precise form is not important. All we will need is that besides the first term, $\tfrac{1}{2}\pi p_{1}^{2}$, each term is at least linear in the coordinate $\psi_{1}$.

Consider the ideal generated by the single coordinate $\Ical = \left< \psi_{1} \right>$. This ideal is automatically closed under the Q-structure since 
\begin{equation}
    \begin{split}
        (\Theta'_{\widetilde{\Mcal}},\psi_{1}) &= (\tfrac{1}{2}\pi p_{1}^{2} + \psi_{1} f(x_{0},\psi_{1},\chi_{0},p_{1}), \psi_{1})
        \\&= 
        (\psi_{1} f(x_{0},\psi_{1},\chi_{0},p_{1}), \psi_{1}) \\
        &\propto \psi_{1} (f,\psi_{1}) \\
        &\in \Ical
    \end{split}
\end{equation}
Performing the coisotropic reduction with respect to this ideal we obtain the structure sheaf
\begin{equation}
    N(\Ical)/\Ical \sim \{ x_{0},p_{1} \} \quad \Rightarrow \quad \Wcal = T^{*}[1]M
\end{equation}
That is, we reproduce the non-commutative manifold $\Wcal$. Further, the symplectic form and Hamiltonian function are easily shown to be the following:
\begin{align}
    \omega_{\Wcal} &= \dd p_{1}\,\dd x_{0}\,, \\
    \Theta_{\Wcal} &= \tfrac{1}{2} \pi p_{1}^{2}\,,
\end{align}
We see then that we precisely reproduce the QP manifold associated to the cotangent lift of the Poisson bivector that we described at the beginning of this section. This construction provides new relations between the Courant sigma model and the Poisson sigma model that is different from the WZW-Poisson model \cite{Klimcik:2001vg}. There they view the Poisson model arising at the boundary of a topological WZW-like theory. Instead, our construction is closer to dimensional reduction and can be viewed as the geometric counterpart of the Courant sigma model reduction found in \cite{Cabrera2022DimensionalRO}.

\section{Examples -- $X = Y$}\label{sec:examples_X=Y}

We will now generalise the previous two sections to allow for cases where the source manifold $X$ wraps the target space fibre $Y$. In particular, we will be interested in the case where $X=Y$. We will see that the reduction procedure requires us to choose some self-wrapping map $\wrap:X\rightarrow X$. The examples we choose are physically motivated and fill our understanding of how brane dualities in M-theory/IIA arise in the QP setting. In particular, when $X=S^{1}$, we will see that our procedure produces the known relations from M-theory/type IIA duality. We will also see that this procedure reproduces other interesting relations between the M5 brane and the heterotic string \cite{Cherkis:1997bx,Park:2009me}.

\subsection{M2 on $S^{1}$}
\label{sec:m2OnS1}

Our first example will be wrapping the M2 brane on an $S^{1}$. This will be very similar to the $n$-brane example in section \ref{sec:n-brane}, except in this case the wrapping will allow for more interesting Hamiltonian functions to be produced.

We start with the QP manifold $\Mcal$ associated to the M2 brane and a source manifold $X$:
\begin{equation}
    \Mcal = T^{*}[3]T[1](N\times S^{1}) \qquad X = S^{1}
\end{equation}
Writing $\Ncal = T^{*}[3]T[1]N$, $\Ycal = T^{*}[3]T[1]S^{1}$, and $\Xcal = T[1]S^{1}$, we will introduce the coordinates
\begin{equation}
    \arraycolsep = 5pt
    \begin{array}{ccc}
        \Ncal & \qquad & \Ycal \\
        \begin{array}{r|cccc}
            \text{coord} & x^{\mu} & \psi^{\mu} & \chi_{\mu} & p_{\mu}  \\ \hline
            \text{deg} & 0 & 1 & 2 & 3
        \end{array} & & \begin{array}{r|cccc}
            \text{coord} & y & \xi & \phi & q  \\ \hline
            \text{deg} & 0 & 1 & 2 & 3
        \end{array}
    \end{array}
\end{equation}
and use coordinates $(\sigma,\dd\sigma)$ on $\Xcal$. The Hamiltonian function and  symplectic form are
\begin{align}\label{eq:omegaMcalM2toF1}
    \omega_{\Mcal} &= \dd p \, \dd x + \dd q\, \dd y - \dd\psi\,\dd\chi - \dd\xi\,\dd\phi \\ \label{eq:ThetaMcalM2toF1}
    \Theta_{\Mcal} &= -\psi p - \xi q + \tfrac{1}{4!}F_{4}\psi^{4} + \tfrac{1}{3!}H_{3}\psi^{3}\xi
\end{align}
We require $F_{4}$ to be $\dr_{S^1}$-closed; we will also use the fact that $H_3 \xi$ is $\dr_{S^1}$-closed (which is automatic).

We transgress this to the mapping space $\Mcal^{\Xcal}$ and choose an ideal whose vanishing locus describes, in degree 0, some embedding $\imath :N \hookrightarrow \Mcal^{\Xcal}$. As explained in section \ref{sec:wrapping_QP}, this depends on the choice of some wrapping map $\wrap:S^{1}\rightarrow S^{1}$. In fact, as stated, the final result only depends on the homotopy class of $w$ and hence we can take, for some $\winding\in \bbZ$,
\begin{equation}
\arraycolsep = 3pt
    \begin{array}{rcl}
        \wrap: S^{1} & \longrightarrow & S^{1}  \\
        \sigma & \longmapsto & \winding\sigma
    \end{array}
\end{equation}
To restrict to this wrapping sector of $\Mcal^{\Xcal}$, we define a coisotropic ideal $\Ical= \left< \Ical_{\Ncal},\Ical_{\Ycal}\right>$ with
\begin{equation}
    \Ical_{\Ycal} = \left< \bm{y}-\winding\sigma, \bm{\xi} + \winding\,\dd \sigma \right>
\end{equation}
As explained in section \ref{sec:wrapping_QP}, this is coisotropic and closed under the Q-structure $D=[\Theta_{\Mcal^{\Xcal}},\cdot]$. The ideal $\Ical_{\Ncal}$ restricts all maps into $\Ncal$ to closed maps. That is, we take
\begin{equation}
    \Ical_\Ncal \supset \left< P_{\co}x_{0},\,P_{\co}\psi_{0},\, P_{\co}\chi_{0},\, P_{\co}p_{0} \right>
\end{equation}
For any coordinate $z^{A}_{k}$ with $\deg z^{A} - k \leq 0$, we need to further restrict to exact maps by including the harmonic representative in the ideal (except for $x_{0}$). Hence, we have
\begin{equation}
    \Ical_{\Ncal} = \left< P_{\co}x_{0},\,P_{\co}\psi_{0},\, P_{\co}\chi_{0},\, P_{\co}p_{0},\, P_{\Hcal}x_{1},\, P_{\Hcal}\psi_{1} \right>
\end{equation}
The ideal $\Ical=\left<\Ical_{\Ncal},\Ical_{\Ycal}\right>$ is clearly coisotropic.

We need to check that $\Ical_{\Ncal}$ is closed under $D = [\Theta_{\Mcal^{\Xcal}},\cdot]$. The transgressed Hamiltonian is
\begin{equation}
    \begin{split}
        \Theta_{\Mcal^{\Xcal}} &= -\int_{\Xcal}\bm{\Theta}_{\Mcal} - \int_{\Xcal} \imath_{\dd}\bm{\vartheta}_{\Mcal} \\
        &= -\int_{\Xcal} - \bm{\psi p} - \bm{\xi q} + \tfrac{1}{4!}F_{4}(\bm{x},\bm{y})\bm{\psi}^{4} + \tfrac{1}{3!}H_{3}(\bm{x},\bm{y}) \bm{\psi}^{3} \bm{\xi} \\
        & \qquad \qquad \qquad \qquad - \int_{\Xcal} \bm{p}\dd \bm{x} + \bm{q}\dd\bm{y} - \tfrac{1}{3}(\bm{\psi}\dd\bm{\chi} + 2\bm{\chi}\dd\bm{\psi} + \bm{\xi}\dd\bm{\phi} + 2\bm{\phi}\dd\bm{\xi})
    \end{split}
\end{equation}
As in the previous cases, the only non-trivial constraint comes from the Poisson bracket between the first term and the harmonic generators of $\Ical_{\Ncal}$. We calculate
\begin{equation}
    (\Theta_{\Mcal}, x) = \psi \,, \qquad \qquad (\Theta_{\Mcal},\psi) =0
\end{equation}
and hence we have
\begin{align}
    \left[ \Theta_{\Mcal^{\Xcal}}, \int_{\Xcal} \bm{\psi}\epsilon \right] &\propto \int_{\Xcal} \bm{(\Theta_{\Mcal},\psi)_{\Mcal}} \,\epsilon = 0 \\
    \left[ \Theta_{\Mcal^{\Xcal}}, \int_{\Xcal} \bm{x} \epsilon \right] & \propto \int_{\Xcal} \bm{(\Theta_{\Mcal},x)_{\Mcal}}\, \epsilon = \int_{\Xcal}\bm{\psi} \epsilon
\end{align}
Evaluating this on $\Ccal$, we take $\psi_{1}$ to be exact and so the integral vanishes when integrated over a constant $\epsilon = \epsilon_{0}$. This shows that the Poisson brackets with the harmonic generators $x_{1}, \psi_{1}$ vanish when evaluated on $\Ccal$, i.e. they are in $\Ical$.

To perform the coisotropic reduction we find the normaliser is generated by
\begin{equation}
    N(\Ical) \sim \{ P_{\Hcal}x_{0},P_{\Hcal}\psi_{0},\, P_{\Hcal}\chi_{1},\, P_{\Hcal}p_{1} ,\, \Ical \}
\end{equation}
and hence the structure sheaf is generated by by
\begin{equation}
\label{eq:CinftyWcalM2toF1}
    C^{\infty}(\Wcal) = N(\Ical)/\Ical \sim \{ x_{0}, \, \psi_{0},\, \chi_{1},\, p_{1} \} \quad \Rightarrow \quad \Wcal =T^{*}[2]T[1]N
\end{equation}
where the coordinates represent harmonic maps. The symplectic form can be derived from the Poisson brackets on $\Mcal^{\Xcal}$, as in section \ref{sec:n-brane}, and we find
\begin{equation}
    \omega_{\Wcal} = -\dd p_{1} \, \dd x_{0} - \dd \psi_{0}\, \dd\chi_{1}
\end{equation}
The Hamiltonian function is given by
\begin{equation}
    \begin{split}
        \Theta_{\Wcal} &= \Pi(\Theta_{\Mcal^{\Xcal}}) = \Pi\left(-\int_{\Xcal}\bm{\Theta}_{\Mcal} - \int_{\Xcal} \imath_{\dd}\bm{\vartheta}_{\Mcal} \right)
    \end{split}
\end{equation}
The second term vanishes when evaluated on harmonic maps where $\dd$ annihilates the maps, except for the term $\bm{q}\dd\bm{y}$. We also get a piece $-\bm{\xi q}$ from the first term. We find
\begin{equation}
    \begin{split}
        \Pi\left( \int_{\Xcal} \bm{\xi q} - \bm{q}\dd\bm{y} \right) &= \int_{\Xcal} -\dd \bm{y}\, \bm{q} - \bm{q}\dd\bm{y} = 0
    \end{split}
\end{equation}
where we pick up a minus sign from commuting $\dd \bm{y}$ (degree 1) through $\bm{q}$ (degree 3). This verifies the statement made in section \ref{sec:wrapping_QP} about this cancellation. We then have
\begin{equation}
    \begin{split}
        \Theta_{\Wcal} &= \Pi\left( \int_{T[1]S^{1}} \bm{\psi p} - \tfrac{1}{4!}F_{4}(\bm{x},\bm{y})\bm{\psi}^{4} - \tfrac{1}{3!} H_{3}(\bm{x},\bm{y})\bm{\psi}^{3}\bm{\xi} \right) \\
        &= \int_{\Xcal} \psi_{0} p_{1} + -\tfrac{1}{4!}F_{4}(x_{0},\winding\sigma)\psi_{0}^{4} +\tfrac{1}{3!}H_{3}(x_{0},\winding\sigma) \psi_{0}^{3}\winding\,\dd\sigma \\
        &= \frac{1}{2\pi}\int_{X} \left(  \psi_{0} p_{1} + \tfrac{\winding}{3!}H_{3}(x_{0},\winding\sigma) \psi_{0}^{3} \right) \dd\sigma \\
        &= \psi_{0} p_{1} + \tfrac{\winding}{3!}\tilde{H}_{3} \psi_{0}^{3}
    \end{split}
\end{equation}
where $\tilde{H}_{3}$ is the average of $H_{3}$ over the fibre.

Under the change of coordinates $p_{1} \to - p_{1}$, we see that we recover the QP manifold associated to the F1 string with $\winding$ units of $\tilde{H}_{3}$ flux, as we would expect from our intuition of M-theory/IIA duality. Note that in the case that $\winding=0$, the physical interpretation seems to break down - we find a string which doesn't couple to the NS 3-form. However, as is noted in \cite{Townsend:1996xj}, this zero winding case corresponds to a scenario in which the original worldvolume is ``collapsed''. This means that the map from the worldvolume to the target space is not an embedding. From the IIA perspective, the resulting string is tensionless and thus the M2 brane must somehow be tensionless. We should discard that case on account of such objects appear not to exist on physical grounds; nevertheless, the QP procedure is well defined.

\subsection{M5 on $S^{1}$}\label{sec:M5_S1}

The next case of interest is wrapping the M5 QP manifold on a circle. The M5 QP manifold was written down in \cite{Arvanitakis:2018cyo} and our expectation is that we should recover that of the D4 brane \cite{Arvanitakis:2022fvv}. We start with the following manifolds.
\begin{equation}
    \Mcal = T^{*}[6]T[1](N\times S^{1})\times \bbR[3] \qquad X = S^{1}
\end{equation}
Writing $\Ncal = T^{*}[6]T[1]N\times \bbR[3]$, $\Ycal = T^{*}[6]T[1]S^{1}$ and $\Xcal = T[1]S^{1}$, we introduce the homogeneous coordinates
\begin{equation}
    \arraycolsep = 5pt
    \begin{array}{ccc}
        \Ncal & \qquad & \Ycal \\
        \begin{array}{r|ccccc}
            \text{coord} & x^{\mu} & \psi^{\mu}& \zeta & \chi_{\mu} & p_{\mu}  \\ \hline
            \text{deg} & 0 & 1 & 3 & 5 & 6
        \end{array} & & \begin{array}{r|cccc}
            \text{coord} & y & \xi & \phi & q  \\ \hline
            \text{deg} & 0 & 1 & 5 & 6
        \end{array}
    \end{array}
\end{equation}
and use coordinates $(\sigma, \dd\sigma)$ for $\Xcal$. We write the symplectic form and Hamiltonian function as
\begin{align}
    \omega_{\Mcal} &= \dd p\, \dd x + \dd q\, \dd y - \dd \psi\, \dd\chi - \dd \xi\, \dd \phi - \tfrac{1}{2}\dd\zeta \,\dd\zeta \\
    \Theta_{\Mcal} &= -\psi p - \xi q + \tfrac{1}{7!}(H_{7}+A\wedge F_{6}) \psi^{7} + \tfrac{1}{6!}F_{6}\psi^{6}\xi + \tfrac{1}{4!}(F_{4} - A\wedge H_{3})\psi^{4}\zeta + \tfrac{1}{3!}H_{3}\psi^{3}\xi \zeta
\end{align}
We included, in this example, a non-trivial connection on the fibre bundle $N\times S^{1}$ which we will assume to be $S^{1}$ invariant. As previously, we can interpret the coefficients to be elements of $\Omega^{i}(N)\times \Omega^{j}(S^{1})$ and we require that they are closed under the $\dd_{S^{1}}$ on $S^{1}$.

We then transgress the structure to the mapping space $\Mcal^{\Xcal}$ and aim to define a suitable ideal $\Ical = \left<\Ical_{\Ncal},\Ical_{\Ycal} \right>$ to perform the coisotropic reduction with respect to. The ideal $\Ical_{\Ycal}$ is taken as in the previous section
\begin{equation}
    \Ical_{\Ycal} = \left<\bm{y}- \winding\sigma,\, \bm{\xi} + \winding \,\dd\sigma \right>
\end{equation}
The ideal $\Ical_{\Ncal}$ is also taken as in the previous section, but now with the additional constraints on the $\bm{\zeta}$ coordinates, restricting them to closed maps. That is, we take
\begin{equation}
    \Ical_{\Ncal} = \left< P_{\co}x_{0},\,P_{\co}\psi_{0},\, P_{\co}\zeta_{0},\, P_{\co}\chi_{0},\, P_{\co}p_{0},\, P_{\Hcal}x_{1},\, P_{\Hcal}\psi_{1} \right>
\end{equation}
Since we have only added co-exact generators to the ideal, the proof of coisotropy and closure under $D$ goes exactly as in the previous case.

Performing the coisotropic reduction, we find the structure sheaf is generated by
\begin{equation}
    C^{\infty}(\Wcal) = N(\Ical)/\Ical \sim \{ x_{0},\, \psi_{0},\, \zeta_{0},\, \zeta_{1},\, \chi_{1},\, p_{1} \}
\end{equation}
which gives
\begin{equation}
    \Wcal = T^{*}[5]T[1]N \times \bbR[2]\times \bbR[3]
\end{equation}
To find the symplectic form, we use the Poisson brackets on $\Mcal^{\Xcal}$ given by \eqref{eq:transgressed_PB} with appropriate insertions of harmonic test functions $\epsilon, \eta$, and find
\begin{equation}
    \omega_{\Wcal} = -\dd p_{1}\, \dd x_{0} - \dd\psi_{0}\, \dd \chi_{1} - \dd \zeta_{1} \,\dd\zeta_{0}
\end{equation}
and the Hamiltonian function is given by\footnote{We are using the fact that the $\imath_{\dd}\bm{\vartheta}_{\Mcal}$ term vanishes, apart from the $\bm{q}\dd\bm{y}$ term, which cancels against the $\bm{\xi q}$ term in $\bm{\Theta}_{\Mcal}$.}
\begin{equation}
    \begin{split}
        \Theta_{\Wcal} &= \Pi(\Theta_{\Mcal^{\Xcal}}) \\
        &= \Pi \left( \int_{T[1]S^{1}} \bm{\psi p} - \tfrac{1}{7!}(H_{7}+A\wedge F_{6}) \bm{\psi}^{7} - \tfrac{1}{6!}F_{6}\bm{\psi}^{6}\bm{\xi} - \tfrac{1}{4!}(F_{4} - A\wedge H_{3})\bm{\psi}^{4}\bm{\zeta} - \tfrac{1}{3!}H_{3}\bm{\psi}^{3}\bm{\xi\zeta}  \right) \\
        &=  \psi_{0}p_{1} + \tfrac{\winding}{6!}\tilde{F}_{6}\psi^{6}_{0} - \tfrac{1}{4!}(\tilde{F}_{4}-A\wedge \tilde{H}_{3})\psi^{4}_{0} \zeta_{1} - \tfrac{\winding}{3!}\tilde{H}_{3}\psi_{0}^{3}\zeta_{0} 
    \end{split}
\end{equation}
where the tilde denotes the average over the $S^{1}$ fibre. For $\winding\neq 0$, we perform a canonical transformation generated by the function $-\tfrac{1}{2\winding}A\psi \zeta_{1}^{2}$ to obtain the Hamiltonian
\begin{equation}
    \Theta_{\Wcal} = \psi_{0} p_{1} + \tfrac{1}{4\winding}F_{2}\psi_{0}^{2}\zeta_{1}^{2} - \tfrac{\winding}{3!}\tilde{H}_{3}\psi_{0}^{3}\zeta_{0} - \tfrac{1}{4!}\tilde{F}_{4}\psi_{0}^{4}\zeta_{1} + \tfrac{\winding}{6!}\tilde{F}_{6}\psi_{0}^{6}
\end{equation}
where $F_{2} = \dd A$. Making the change of coordinates $p_{1} \to - p_{1}$, $\zeta_{i} \to - \zeta_{i}$ puts the QP manifold in the canonical form of that associated to the D4 brane \cite{Arvanitakis:2022fvv}.

\subsection{M5 on $X_{4}$}

The next example will be to wrap the M5 brane over a 4-manifold $X_{4}$. In \cite{Cherkis:1997bx,Park:2009me} it was shown that one could reproduce the non-critical heterotic string through such a reduction, where the dimension of the gauge group was related to the cohomology of the wrapping manifold. We will start with the manifolds
\begin{equation}
    \Mcal = T^{*}[6]T[1](N\times X_{4}) \times \bbR[3] \qquad X = X_{4}
\end{equation}
Writing $\Ncal = T^{*}[6]T[1]N\times \bbR[3]$, $\Ycal = T^{*}[6]T[1] X_{4}$, $\Xcal = T[1]X_{4}$, we introduce the homogeneous coordinates as in the previous section
\begin{equation}
    \arraycolsep = 5pt
    \begin{array}{ccc}
        \Ncal & \qquad & \Ycal \\
        \begin{array}{r|ccccc}
            \text{coord} & x^{\mu} & \psi^{\mu}& \zeta & \chi_{\mu} & p_{\mu}  \\ \hline
            \text{deg} & 0 & 1 & 3 & 5 & 6
        \end{array} & & \begin{array}{r|cccc}
            \text{coord} & y^{m} & \xi^{m} & \phi_{m} & q_{m}  \\ \hline
            \text{deg} & 0 & 1 & 5 & 6
        \end{array}
    \end{array}
\end{equation}
where now $\alpha = 1,...,4$, and we use the DG coordinates $(\sigma^{\alpha},\dd\sigma^{\alpha})$ on $\Xcal$. In these coordinates the symplectic form and Hamiltonian function take the form
\begin{align}
    \omega_{\Mcal} &= \dd p\, \dd x + \dd q\, \dd y - \dd \psi\, \dd\chi - \dd \xi\, \dd \phi - \tfrac{1}{2}\dd\zeta \,\dd\zeta \\
    \begin{split}
    \Theta_{\Mcal} &= - \psi p - \xi q + \tfrac{1}{7!}H_{7}\psi^{7} + \tfrac{1}{6!}H_{6} \psi^{6}\xi + \tfrac{1}{2}\tfrac{1}{5!}H_{5}\psi^{5}\xi^{2} + \tfrac{1}{3!}\tfrac{1}{4!}H_{4}\psi^{4}\xi^{3} + \tfrac{1}{4!}\tfrac{1}{3!}H_{3}\psi^{3}\xi^{4} \\
    & \qquad + \tfrac{1}{4!}F_{4}\psi^{4}\zeta + \tfrac{1}{3!}F_{3}\psi^{3}\xi \zeta + \tfrac{1}{2}\tfrac{1}{2}F_{2}\psi^{2}\xi^{2}\zeta + \tfrac{1}{3!}F_{1}\psi\xi^{3}\zeta + \tfrac{1}{4!}F_{0} \xi^{4}\zeta
    \end{split}
\end{align}
where we have taken a trivial connection on the $X_{4}$ bundle again. As before, we can view the coefficients as differential forms on $Y$ valued in $\Omega^{k}(N)$ that we take to be $\dr_Y$-closed.

We transgress this structure to $\Mcal^{\Xcal}$ and define a coisotropic ideal $\Ical= \left<\Ical_{\Ncal},\Ical_{\Ycal}\right>$. To define the ideal $\Ical_{\Ycal}$ we need to choose some wrapping map $\wrap:X_{4}\to X_{4}$. Restriction to this winding sector of $\Mcal^{\Xcal}$ is given by
\begin{equation}
    \Ical_{\Ycal} = \left< \bm{y}-\wrap,\, \bm{\xi} + \dd \wrap \right>
\end{equation}
It is easy to verify that this is coisotropic and closed under $D$. The ideal $\Ical_{\Ncal}$ is similar to that for the circle reduction done in the previous section, except now our transgressed coordinates are $k$-forms\footnote{As noted in section \ref{sec:wrapping_QP}, the transgressed coordinates $z^{A}_{k}$ for $k\geq 2$ should be viewed as differential forms evaluated in some affine bundle. Our construction is still well-defined so for simplicity we will ignore this subtlety here.} for $k=0,...,4$. This means that we need to include more co-exact generators and harmonic generators to remove unwanted coordinates. We take
\begin{equation}
    \Ical_{\Ncal} = \left< P_{\co}x_{k},\, P_{\co}\psi_{k},\, P_{\co}\zeta_{k},\, P_{\co} \chi_{k},\, P_{\co} p_{k},\, P_{\Hcal}x_{i},\, P_{\Hcal}\psi_{i},\,  P_{\Hcal} \zeta_{j} \,|\, i > 0,\, j >2 \right>
\end{equation}
We need to check whether this is closed under $D$. As in previous cases, the only non-trivial checks come from the harmonic generators. The Q-structure $D$ acting on the harmonic generators $x_{i},\psi_{i}$ return an element of $\Ical_{\Ncal}$ precisely as in previous cases so we need only check the closure of $D\zeta_{3},\, D\zeta_{4}$. Once again, this can be done by calculating
\begin{equation}
    (\Theta_{\Mcal},\zeta)_{\Mcal} = \tfrac{1}{4!}F_{4} \psi^{4} + \tfrac{1}{3!}F_{3}\psi^{3}\xi + \tfrac{1}{4}F_{2}\psi^{2}\xi^{2} + \tfrac{1}{3!}F_{1}\psi\xi^{3} + \tfrac{1}{4!}F_{0}\xi^{4}
\end{equation}
We then transgress this function to $\Mcal^{\Xcal}$ and evaluate it on the vanishing locus $\Ccal$ of $\Ical$. We then check whether the following vanishes
\begin{equation}
    \int_{\Xcal} \bm{(\Theta_{\Mcal},\zeta)_{\Mcal}} \, \epsilon
\end{equation}
for suitable harmonic test functions $\epsilon$. To determine the conditions coming from $\zeta_{4}$, we take $\epsilon= \epsilon_{0}$ a constant function. We then get the constraint
\begin{equation}
    \int_{X} \wrap^{*}(F_{0})\, \epsilon_{0} \overset{!}{=} 0
\end{equation}
where we are using the fact that $F_{0}$ is a 4-form on $Y$ which we pull back to $X$ via the wrapping map. Similarly, the conditions coming from $\zeta_{3}$ are given by choosing an arbitrary harmonic 1-form $\epsilon = \epsilon_{1}$
\begin{equation}\label{eq:M5_X4_zeta3_constraint}.
    \int_{X} \wrap^{*}(F_{1})\wedge \epsilon_{1} \overset{!}{=} 0
\end{equation}
This puts constraints on the coefficients $F_{0}, F_{1}$ which can be most easily satisfied if they vanish, i.e. they act as obstructions to the reduction. Note that in some cases, e.g. for $X=\text{K3}$, there are no non-trivial harmonic 1-forms and so \eqref{eq:M5_X4_zeta3_constraint} gives no constraints.

Assuming these constraints are satisfied, the coisotropic reduction with respect to $\Ical = \left<\Ical_{\Ncal},\Ical_{\Ycal}\right>$ gives that the structure sheaf is generated by
\begin{equation}
    C^{\infty}(\Wcal) = N(\Ical)/\Ical \sim \{ x_{0},\, \psi_{0},\, \zeta^{a}_{2},\, \chi_{4},\, p_{4} \} \quad \Rightarrow \quad \Wcal = T^{*}[2]T[1]N \times H^{2}(X)[1]
\end{equation}
We have introduced an index $a$ parameterising a basis $\{e_{a}\}$ of $\mathcal{H}^{2}(X_{4})$, and have expanded $\zeta_{2} \in \mathcal{H}^{2}$ as $\zeta_{2}^{a} e_{a}$. Using the Poisson brackets on $\Mcal^{\Xcal}$, we get the symplectic form and the Hamiltonian function on $\Wcal$ to be
\begin{align}
    \omega_{\Wcal} &= \dd p_{4}\, \dd x_{0} - \dd \psi_{0} \, \dd \chi_{4} - \tfrac{1}{2} \kappa_{ab}\dd \zeta^{a}_{2}\, \dd\zeta^{b}_{2} \\
    \Theta_{\Wcal} &= \Pi(\Theta_{\Mcal^{\Xcal}}) = -\psi_{0} p_{4} + \tfrac{1}{3!}\tilde{H}_{3}\psi^{3} + \tfrac{1}{2}\tilde{F}_{a}\psi^{2}\zeta_{2}^{a}
\end{align}
where
\begin{equation}
    \tilde{H}_{3} = \int_{X}\wrap^{*}(H_{3})\, , \qquad \tilde{F}_{a} = \int_{X}\wrap(F_{2})\wedge e_{a}\, , \qquad \kappa_{ab} = \int_{X} e_{a}\wedge e_{b}
\end{equation}

We get the canonical form of the QP manifold associated to a heterotic string with abelian gauge group of dimension $b_{2}(X_{4})$. The Killing form on the gauge group is also given by the symmetric form $\kappa_{ab}$ on $H^{2}(X_{4})$. For example, if $X_{4} = T^{4}$, we get an abelian gauge group of dimension $b_{2}(T^{4})$ with Killing form of signature $(3,3)$. If $X_{4}=\text{K3}$, then we get a gauge group of dimension $b_{2}(\text{K3}) = 22$ with Killing form of signature $(3,19)$. This matches the results of \cite{Cherkis:1997bx,Park:2009me}. The fact that we can only obtain abelian gauge groups arises because we are assuming that we are reducing on smooth manifolds. Degenerations of $X_{4}$ to some singular space should lead to gauge enhancement and non-abelian groups.

\section{AKSZ sigma models and brane wrapping}\label{sec:lifting_AKSZ}
In previous sections we obtained an NQP manifold $\mathcal W$ from a coisotropic reduction of the mapping space $\Mcal^{\Xcal}$ with respect to a coisotropic submanifold $\mathcal C$ that is invariant with respect to the Q-structure $D=Q_{\Mcal}+\dr_{\Xcal}$ on $\Mcal^{\Xcal}$. (In the expression for $D$ we have the lifts of vector fields on the target and source to the mapping space.) We will now point out that these data give rise --- essentially trivially! --- to a reduction of AKSZ sigma models from an AKSZ model with target $\Mcal$ to an AKSZ model with target $\Wcal$.

We start with the BV manifold of an AKSZ sigma model with target $\Mcal$ where the source takes the form $\Xcal\times \Scal$. The N-manifold $\Scal$ is taken to be $T[1]S$ where the (bosonic) manifold $S$ has dimension $\dim S=n+1 -\dim X$ ($n$ being the degree of the target P-structure). Then the BV master action is the hamiltonian corresponding to the Q-structure on $\Mcal^{\Xcal\times \Scal}$ given by 
\be
Q_{\mathrm{BV}}\equiv Q_{\Mcal}+\dr_{\Xcal\times \Scal}
\ee
where again $Q_{\Mcal}$ denotes the lift to $\Mcal^{\Xcal\times \Scal}$ of the target space $\Mcal$ Q-structure of the same name, and $\dr_{\Xcal\times \Scal}$ is the lift of the source $\Xcal\times \Scal$ de Rham differential again to $\Mcal^{\Xcal\times \Scal}$. Since the source is a product we can write $\dr_{\Xcal\times \Scal}=\dr_\Xcal+\dr_\Scal$.

The key point that leads to reduction is that we can write
\be
\Mcal^{\Xcal\times \Scal}=(\Mcal^\Xcal)^\Scal
\ee
which is known as the product-exponential adjunction. Explicitly, this corresponds to interpreting a function $f\in \Mcal^{\Xcal\times \Scal}$, which is a function $f(x,s)$ of two arguments, as a function $s\to f(\bullet,s)$ where $f(\bullet,s)$ is a function of $x\in \Xcal$ for each $s\in \Scal$.\footnote{The definition of mapping spaces for graded manifolds is such that this property is true; see e.g.~\cite{Roytenberg:2006qz}.} Since $\Mcal^{\Xcal}$ is a QP-manifold and $\Scal$ is an NQ-manifold with an integral measure we can \textbf{consider the BV structure on $\Mcal^{\Xcal\times \Scal}$ as arising from an AKSZ construction with source $\Scal$ and target  $\Mcal^{\Xcal}$}. If $\Ccal$ is coisotropic in $\Mcal^\Xcal$, then the mapping space $\Ccal^\Scal$ will be a coisotropic submanifold in $(\Mcal^\Xcal)^\Scal\cong \Mcal^{\Xcal\times \Scal}$.

The reduced AKSZ sigma model will be given by the coisotropic reduction of $\Mcal^{\Xcal\times \Scal}$ with respect to $\Ccal^\Scal$. We need to confirm that $\Ccal^\Scal$ is invariant with respect to $Q_\mathrm{BV}$, so that the BV master action reduces. We rewrite $Q_\mathrm{BV}$ as
\be
Q_\mathrm{BV}= (Q_\Mcal + \dr_{\Xcal})+\dr_{\Scal}= \hat D + \bar \dr_{\Scal}
\ee
where in the last formula $\hat D$ is the lift from $\Mcal^{\Xcal}$ to $(\Mcal^\Xcal)^\Scal$ of the vector field $D$ on $\Mcal^\Xcal$, while $\bar\dr_{\Scal}$ is the lift of $\dr_{\Scal}$ from $\Scal$ to $(\Mcal^\Xcal)^\Scal$. We denote these lifts explicitly now because it is the properties of these lifts that guarantee the reduction: if $V_\Scal$ is any vector field on $\Scal$, then the lift $\bar{V}_\Scal$ always leaves $\Ccal^\Scal$ invariant (for any submanifold $\Ccal$ of $\Mcal^\Xcal$); $\Ccal^\Scal$ is invariant for $\hat D$ if $\Ccal$ is invariant for $D$. Therefore $Q_\mathrm{BV}$ gives rise to a homological (and hamiltonian) vector field on the coisotropic reduction of $\Mcal^{\Xcal\times \Scal}$, which is simply $\Wcal^{\Scal}$. Using the results of appendix \ref{app:CoisoreductionOfPoisson} we find that the new BV master action is given by evaluating the original action on $\Ccal^\Scal$. In all examples we have investigated the result is another topological field theory of AKSZ type.

In summary, the brane wrapping of QP manifolds that we already discussed  always leads to a brane wrapping procedure that takes the BV master action associated to an AKSZ topological field theory and produces the BV master action of another topological field theory.

\subsection{AKSZ 3-brane to membrane example}
To illustrate, we will treat the reduction of the AKSZ sigma model corresponding to the wrapping of an M2 algebroid on a circle that we discussed in Section \ref{sec:m2OnS1}. This is a reduction of the 4D topological field theory of Ikeda and Uchino \cite{Ikeda:2010vz} to a (3D) Courant sigma model.

This example thus has $\Xcal=T[1]S^1$ and the coisotropic submanifold $\Ccal \subset \Mcal^\Xcal$ is given
by
\be\begin{split}
\dr_{\Xcal} p_0 = 0\,, \quad \dr_{\Xcal} x_0^\mu=0\,,\quad P_{\mathcal H}x_1^\mu=P_{\mathrm{co}}x_1^\mu=0\,,\quad  y_0=w\sigma\,,\quad y_1=0\\
\dr_{\Xcal} \chi_0 = 0 \,, \quad \dr_{\Xcal}\psi_0^\mu=0\,,\quad  P_{\mathcal H}\psi_1^\mu=P_{\mathrm{co}}\psi_1^\mu=0\,,\quad \xi_0=0\,,\quad \xi_1=-w\,.
\end{split}
\ee
We have used the superfield expansion of $\bm{Z}^{\bar A}=\{\bm{x}^\mu,\bm{y},\bm{\psi}^\mu,\bm{\xi},\cdots\}$ in form degree (so $\bm x(\sigma,\dr \sigma)= x_0(\sigma) + x_1(\sigma)\dr \sigma$ etc.)

For the original (4D) AKSZ theory degree-counting to work we set $\Scal=T[1]S$ where $S$ can be any 3-manifold, so that $X\times S=S^1\times S$ is the four-dimensional worldvolume. Using the product-exponential adjunction to write $\Mcal^{\Xcal\times \Scal}\cong (\Mcal^\Xcal)^\Scal$ amounts to promoting the components $Z_k^{\bar A}$ of the superfields $\bm{Z}^{\bar A}$ defining a map $\Mcal^\Xcal$ to superfields $\bm{Z}_k^{\bar A}$ that now depend on the $\Scal$ coordinates $\{s,\dr s\}$ \emph{as well as} the $\Xcal$ coordinates ($\{\sigma,\dr \sigma\}$ in this case). Then the coisotropic submanifold $\Ccal^{\Scal}$ is the locus of functions $\Scal\to \Mcal^{\Xcal}$ such that 
\be\begin{split}
\dr_{\Xcal} \bm{p}_{0} = 0 \,, \quad \dr_{\Xcal} \bm{x}_0^\mu=0\,,\quad  P_{\mathcal H}\bm x_1^\mu=P_{\mathrm{co}}\bm x_1^\mu=0\,,\quad  \bm{y}_0=w\sigma\,,\quad \bm{y}_1=0\\
\dr_{\Xcal}\bm{\chi}_{0} = 0\,, \quad \dr_{\Xcal}\bm{\psi}_0^\mu=0\,,\quad P_{\mathcal H}\bm \psi_1^\mu=P_{\mathrm{co}}\bm \psi_1^\mu=0 \,,\quad \bm{\xi}_0=0\,,\quad \bm{\xi}_1=-w\,.
\end{split}
\ee
where all bolded expressions depend on $\{\sigma,s,\dr s\}$. (The projectors to co-exact/harmonic pieces refer to the Hodge decomposition with respect to $\Xcal$ as above.)

We can explicitly check the claim that $\Ccal^\Scal$ is invariant with respect to $Q_\mathrm{BV}=\hat D+\bar{\dr}_\Scal$. E.g.
\be
\hat D\int_{\Scal\times \Xcal} (\bm y-w\sigma)\epsilon=\int_{\Scal\times \Xcal}(\bm \xi + \dr \sigma \pd_\sigma \bm y)\epsilon \overset{\mod \Ical(\Ccal^\Scal)}{=}\int_{\Scal\times \Xcal} (-w\dr \sigma + \dr \sigma w)\epsilon=0 \ee
(We smeared against $\epsilon \in C^\infty(\Scal\times \Xcal)$ and employed \eqref{eq:closure_of_ideal}). The other differential $\bar{\dr}_\Scal$ leaves the ideal invariant independently.  This way we may confirm explicitly that $S_\text{BV}$ lies in $N(\Ical(\Ccal^\Scal))$.

It remains to calculate the reduced BV master action, which amounts to calculating $\Pi(S_\text{BV})$ where $\Pi$ implements the quotient modulo $\Ical(\Ccal^\Scal)$. $S_\text{BV}$ is the hamiltonian for $Q_\text{BV}= D+\dr_\Scal=Q_\Mcal+\dr_{\Xcal}+\dr_{\Scal}\equiv Q_\Mcal+\dr$ which is explicitly given by formula \eqref{eq:transgressedhamiltonian}, which is a linear combination of $\int_{\Xcal\times \Scal} \bm{\Theta_\Mcal}$ and $
\int_{\Xcal\times \Scal} \iota_\dr \bm \vartheta_\Mcal
$, for $\bm \vartheta_\Mcal$ the transgression of a symplectic potential on $\Mcal$ that satisfies $\dr_\Mcal \vartheta_\Mcal=\omega_\Mcal$, $\omega_\Mcal$ being given in \eqref{eq:omegaMcalM2toF1}.
The bolded quantities are superfields corresponding to $\Mcal^{\Xcal\times \Scal}$ now. We then calculate
\be
\begin{split}
\Pi\int_{\Xcal\times \Scal} \iota_\dr \bm \vartheta_\Mcal=\Pi\int_{\Xcal\times \Scal} \bm{p}\dr \bm{x} + \bm{q}\dr \bm{y} - \bm{\chi}\dr \bm{\psi} - \bm \phi \dr \bm \xi\\=\int_\Scal ({\textstyle \int_\Xcal}\bm{p}_1\dr \sigma) \dr_\Scal \bm x_0 + w ({\textstyle \int_\Xcal} \bm q_0 \dr \sigma ) - ({\textstyle \int_\Xcal} \bm \chi_1 \dr \sigma) \dr_\Scal \bm \psi_0
\end{split}
\ee
Note that terms involving $\bm{x}_{1}$, $\bm{\psi}_{1}$ will generate $\dd_{\Xcal}$-exact terms which will vanish under the $\int_{\Xcal}$ integral. Using \eqref{eq:ThetaMcalM2toF1},
\be
\Pi \int_{\Xcal\times \Scal} \bm{\Theta_\Mcal} =\int_\Scal -\psi_0 ({\textstyle \int_\Xcal} \bm p_1 \dr \sigma) -w ({\textstyle \int_\Xcal}\dr \sigma \bm q_0)+0-w({\textstyle \int_\Xcal} \tfrac{1}{3!} H_3 (\bm \psi_0)^3 \dr \sigma)\,.
\ee
We then read off the sign factors from \eqref{eq:transgressedhamiltonian} to find
\be
\begin{split}
\Pi S_{\text{BV}}=\Pi \Big(-\int_{\Xcal\times \Scal} \bm{\Theta_\Mcal} +\int_{\Xcal\times \Scal} \iota_\dr \bm \vartheta_\Mcal \Big)\\
=\int_\Scal \psi_0 ({\textstyle \int_\Xcal} \bm p_1 \dr \sigma) +w({\textstyle \int_\Xcal} \tfrac{1}{3!} H_3 (\bm \psi_0)^3 \dr \sigma) +({\textstyle \int_\Xcal}\bm{p}_1\dr \sigma) \dr_\Scal \bm x_0 - ({\textstyle \int_\Xcal} \bm \chi_1 \dr \sigma) \dr_\Scal \bm \psi_0 \,.\end{split}
\ee
The signs were such that the terms $w{\textstyle \int_\Xcal} \bm q_0 \dr \sigma$ cancelled.

In the above expression we can identify the integrated expressions $({\textstyle \int_\Xcal}\bm{p}_1\dr \sigma)\,, ({\textstyle \int_\Xcal} \bm \chi_1 \dr \sigma)$ as the conjugate momenta superfields (with degrees 2, 1 respectively) that appear in the Courant sigma model for an exact Courant algebroid structure defined by the 3-form $w H_3$. The result we calculated via coisotropic reduction of the original (4-dimensional) AKSZ topological sigma model is identical to the AKSZ sigma model constructed directly from the wrapped QP manifold $\Wcal$ with source manifold $\Scal$ (see \eqref{eq:CinftyWcalM2toF1}).

Therefore we  have recovered the correct relation between the M-theory fluxes, the M2-brane winding $w$, and the type IIA NS-flux $w H_3$ seen by the fundamental strings that arise as the M-theory circle $X=S^1$ is shrunk to zero, all at the level of the corresponding topological sigma models.

\section{Conclusions}\label{sec:conclusions}

We defined a reduction procedure of NQP manifolds $\Mcal \to \Wcal$ which encompasses the properties of wrapped branes. This is consistent with the AKSZ procedure in the sense that the reduction naturally lifts to a reduction of the AKSZ theory with target $\Mcal$ to the AKSZ theory with target $\Wcal$. We applied this to many examples, including many physically motivated examples of wrapped branes and we saw that it reproduced the known M-theory/IIA dualities. We also were able to find a novel relation between the Courant algebroid and the Poisson algebroid through this reduction. We expect that our work will have many interesting applications to other topological AKSZ theories. 

One can ask how general our procedure is, or whether it is possible to relax some of the assumptions made in section \ref{sec:wrapping_QP}. For example, can we relax the trivial bundle condition $M= N\times Y$, perhaps by introducing some flat connection similar to \cite{Bonechi:2010tbl}? We can also ask whether we can extend our construction to manifolds $X$ with boundary. We can also relax the constraint on $\Xcal = T[1]X$, and instead just take $\Xcal$ to be some DG manifold with some invariant measure of degree $n+1$. For example, we can try to extend the reduction procedure to $\Xcal = T^{1,0}[1]X$ for some complex manifold $X$ with $\dim_{\bbC}X = n+1$. We could then apply the reduction to, say, the work of \cite{Qiu:2009zv}.

In section \ref{sec:Courant->Poisson}, we found an interesting relation between the Courant algebroid and the Poisson algebroid QP structures. This was based on the embedding of the Poisson differential $\dd_{\pi}$ into $T\oplus T^{*}$. There are other interesting differentials that can appear in these Courant algebroids \cite{Ashmore:2021pdm} that are associated to topological theories on $\mathrm{G}_{2}$ and $\Spin(7)$ manifolds. One can try to embed these differentials in the language of QP structures and perform the reduction to get new topological models associated to these special holonomy manifolds. There are also similar structures that appear in higher algebroids. That is, one can define the notion of a Dirac structure for these higher algebroids and define the associated differential \cite{severa2001some,Tennyson:2021qwl,Ashmore:2019rkx,Ashmore:2019qii,10.1093/imrn/rnx163,2021JGP...16104055C,Arvanitakis:2021wkt}. These can be embedded into the Q-structure of the QP manifolds associated to these higher algebroids. Their reductions may provide further insight into supersymmetric geometries of string/M-theory.

\section*{Acknowledgements}
We would like to thank Marco Zambon, Chris Blair, Dan Thompson, Ondrej Hulik, and Maxime Grigoriev for very helpful discussions during this project. The authors are grateful to the Mainz Institute for Theoretical Physics (MITP) of the DFG Cluster of Excellence PRISMA\textsuperscript{*} (Project ID 39083149), for its hospitality and its partial support during the completion of this work. ASA is supported by the FWO-Vlaanderen through the project G006119N, as well as by the Vrije Universiteit Brussel through the Strategic Research Program “High-Energy Physics”. He is also supported by an FWO Senior Postdoctoral Fellowship (number 1265122N). DT is supported by the NSF grant PHYS-2112859. Part of the research for this project was performed while DT was supported by the EPSRC New Horizons Grant ``New geometry from string dualities'' EP/V049089/1.

\appendix 

\section{Notation}\label{app:notation}

\subsubsection*{Commutative Manifolds}
\begin{tabular}{p{0.1\textwidth} p{0.8\textwidth}}
    $M$ & Starting/parent commutative manifold which is always a product manifold of a base and a fibre to be wrapped \\
    $N$ & The commutative manifold which is the base of the trivial fibre bundle $M$ \\
    $Y$ & The fibre of the trivial bundle $M$. This is the manifold over which we wrap the branes. \\
    $X$ & The fibre of the brane that is wrapped over $Y$
\end{tabular}

\subsubsection*{Non-commutative Manifolds}

\begin{tabular}{p{0.1\textwidth} p{0.8\textwidth}}
    $\Mcal$ & Starting/parent QP manifold \\
    $\Ncal$ & A submanifold of $\Mcal$ which is the natural QP manifold restricted to the base of the fibration \\
    $\Ycal$ & A submanifold of $\Mcal$ which is the natural QP manifold restricted to the fibre; usually $\Ycal=T^\star[n]T[1] Y$ \\
    $\Xcal$ & The shifted tangent bundle $T[1]X$; the source of the mapping space $\Mcal^\Xcal$ \\
    $\Wcal$ & Final wrapped QP manifold \\
    $\Mcal^{\Xcal}$ & $\mathrm{maps}(\Xcal \rightarrow \Mcal)$ \\
    $\Scal$ & A DG manifold with invariant measure of degree $n+1-\dim X$
\end{tabular}

\subsubsection*{Indices}

\begin{tabular}{p{0.1\textwidth} p{0.8\textwidth}}
    $A,B,C,...$ & Indices along $\Mcal, \Ncal$ \\
    $\mu,\nu,\rho,...$ & Indices along $N$ \\
    $m,n,p,...$ & Indices along $Y$ \\
    $\alpha,\beta,\gamma,...$ & Indices along $X$ \\
    $r,s,t,...$ & Indices correspnding to degree shifted real lines $\bbR[n_{r}]$ \\
    $a,b,c,...$ & Indices for a basis of differential forms on $X$
\end{tabular}

\subsubsection*{Coordinates}

\begin{tabular}{p{0.1\textwidth} p{0.8\textwidth}}
    $Z^{ A}$ & Homogeneous coordinates on $\Mcal$ \\
    $z^{A}$ & Homogeneous coordinates on $\Ncal$ \\
   $x^{\mu}$ & Degree 0 coordinates on $\Ncal$ \\
   $\psi^{\mu}$ & Degree 1 coordinates on $\Ncal$ parameterising the fibre of $T[1]N$ \\
   $p_{\mu}$ & Coordinate dual to $x^{\mu}$ \\
   $\chi_{\mu}$ & Coordinate dual to $\psi^{\mu}$ \\
   $y^{m}$ & Degree 0 coordinates on $\Ycal$ parameterising the fibre of $T[1]Y$ \\
   $\xi^{m}$ & Degree 1 coordinates on $\Ycal$ \\
   $q_{m}$ & Coordinate dual to $y^{\alpha}$ \\
   $\phi_{m}$ & Coordinate dual to $\xi^{\alpha}$ \\
   $(\sigma^{\alpha}, \dd\sigma^{\alpha})$ & Coordinates for the DG manifold $(\Xcal,\dd)$ such that $\dd(\sigma^\alpha)=\dd \sigma^\alpha$ \\
   $\zeta^{r}$ & Homogeneous coordinates corresponding to degree shifted real lines $\bbR[n_{r}]$ \\
   $\bm{Z}^{A}$ & Transgressed coordinates of $\Mcal^{\Xcal}$ \\
   $Z^{A}_{k}$& An expansion of the transgressed coordinates $\bm{Z}^{A}$ into differential $k$-forms \\
   $Z^{A,a}_{k}$ & A coordinate labelling the harmonic $k$-forms, labelled by $a$, associated to the transgressed coordinate $\bm{Z}^{A}$ \\
\end{tabular}

\subsubsection*{Functions and differential forms}

\begin{tabular}{p{0.1\textwidth} p{0.8\textwidth}}
    $\Omega^{k}$ & The space of differential $k$-forms \\
    $\mathcal{H}^{k}$ & Harmonic $k$-forms \\
    $e_{k,a}$ & A basis of harmonic $k$-form(s) (occasionally the $k$ is dropped) \\
    $\Theta_{\Mcal}$ & The Hamiltonian function of $\Mcal$ (similarly for $\Ncal,\Wcal,...$) \\
    $\omega_{\Mcal}$ & The symplectic form of $\Mcal$ (similarly for $\Ncal,\Wcal,...$) \\
    $\vartheta_{\Mcal}$ & The canonical symplectic potential of $\Mcal$ (similarly for $\Ncal,\Wcal,...$) \\
    $(\cdot,\cdot)_{\Mcal}$ & The Poisson bracket for $\Mcal$ (similarly for $\Ncal,\Wcal,...$) \\
    $[\cdot,\cdot]$ & The Poisson bracket on $\Mcal^{\Xcal}$
\end{tabular}

\subsubsection*{Miscellaneous}
\begin{tabular}{p{0.1\textwidth} p{0.8\textwidth}}
    $\wrap$ & Wrapping map $X\to Y$ \\
    $\winding$ & Winding number/matrix of a circle/torus over itself \\
    $\Ical$ & The coisotropic ideal within $\Mcal^{\Xcal}$ \\
    $\Ccal$ & The vanishing locus of $\Ical$ within $\Mcal^{\Xcal}$
\end{tabular}

\begin{tabular}{p{0.05\textwidth} p{0.85\textwidth}}
    
\end{tabular}

\section{QP manifolds}\label{app:QP_manifolds}

\subsubsection*{Graded manifolds}

A graded manifold $\Mcal$ is a supermanifold whose coordinates come equipped with a $\bbZ$ grading.\footnote{From \cite{JOZWIKOWSKI2016212}, the consistency of the $\bbZ$ grading of coordinates comes from the existence of a global degree counting vector field $\varepsilon$ and transition functions which preserve degree.} One can always find homogeneous coordinates $Z^{A}$ of definite degree, where $\deg Z^{A} \, \mathrm{mod} \, 2$ is the Grassman parity of the coordinate. We will denote by $A$ the degree of $Z^{A}$ and so we have
\begin{equation}
    Z^{A}Z^{B} = (-1)^{AB}Z^{B}Z^{A}
\end{equation}
The sheaf of functions on $\Mcal$ splits into subsheafs $C^{\infty}_{n}(\Mcal)$ of functions of definite degree. The degree of a homogeneous function $f$ is measured by the degree counting vector field $\varepsilon$ (The `Euler vector field') via
\begin{equation}
    \varepsilon (f) = \deg(f) f
\end{equation}
In local homogeneous coordinates $Z^{A}$, we have
\begin{equation}
    \varepsilon = \sum_{A}\deg(Z^{A})Z^{A} \frac{\partial}{\partial Z^{A}}
\end{equation}

Unless otherwise stated, all derivations are left derivations. Hence, the de Rham $\dd$ is
\begin{equation}
    \dd f = \dd Z^{A} \partial_{A} f
\end{equation}
and any homogeneous (in degree) vector field $X$ acts as
\begin{equation}
    X(fg) = X(f)g + (-1)^{Xf} f X(g)
\end{equation}
where we have used the shorthand $X,f$ for the degree of the respective components. In local coordinates we can write $X = X(Z)^{A}\partial_{A}$, and so $\deg X = \deg X^{A} - \deg Z^{A}$. We also define
\begin{equation}
    \deg (\dd f) = \deg f + 1
\end{equation}
For this to be consistent with $\imath_{A} \dd Z^{B} = \delta^{B}{}_{A}$, where $\imath_{A}$ denotes contraction with the vector field $\partial_{A}$, we require that the interior product has degree
\begin{equation}
    \deg \imath = -1
\end{equation}

\subsubsection*{Poisson and symplectic structures}

A graded Poisson structure of degree $-n$ is defined to satisfy
\begin{equation}
    (f,g) = (-1)^{1+(f+n)(g+n)}(g,f)
\end{equation}
and the graded Jacobi identity
\begin{equation}
    (f,(g,h)) = ((f,g),h) + (-1)^{(f+n)(g+n)}(g,(f,h))
\end{equation}
for all homogeneous functions $f,g,h$. It also acts as a left derivation on the right hand arguments, but a right derivation on the left hand arguments. That is
\begin{equation}\label{eq:Poisson_LR_derivation}
    \begin{split}
        (f,gh) &= (f,g)h + (-1)^{(f+n)g}g(f,h) \\
        (fg,h) &= f(g,h) + (-1)^{(h+n)g}(f,h)g
    \end{split}
\end{equation}
If the Poisson structure is induced from a symplectic structure $\omega$, we have that
\begin{equation}\label{eq:poisson_symplectic_relation}
    \imath_{X_{f}} = (-1)^{f} \dd f \qquad X_{f} := (f,\cdot)
\end{equation}

In local homogeneous coordinates we can write
\begin{equation}
    \omega = \tfrac{1}{2}\dd Z^{A} \omega_{AB} \dd Z^{B}
\end{equation}
which implies the symmetry
\begin{equation}
    \omega_{AB} = (-1)^{1+AB+n(A+B)}\omega_{BA}
\end{equation}
If we define $\omega^{AB}$ via $\omega^{AB}\omega_{BC} = \delta^{A}{}_{C}$, then \eqref{eq:poisson_symplectic_relation} implies
\begin{equation}
    (f,g) = (-1)^{f}\partial^{R}_{A}f\, \omega^{AB}\,\partial_{B} g
\end{equation}
where $\partial^{R}_{A}$ is defined by $\dd f = \dd Z^{A} \partial_{A} f  = \partial^{R}_{A} \, \dd Z^{A}$. Note that it is not a right derivation by itself, but the combination $(-1)^{f}\partial^{R}_{A}f $ is a right derivation. This is consistent with \eqref{eq:Poisson_LR_derivation}. Note that this implies
\begin{equation}
    (Z^{A},Z^{B}) = (-1)^{A}\omega^{AB}
\end{equation}

The symplectic potential is defined such that $\dd\vartheta = \omega$, and can be defined canonically through the Euler vector field $\varepsilon$. We have that\footnote{More generally, the Lie derivative on any graded differential form along a vector field $X$ is given by $\mathcal{L}_{X} = \imath_{X} \dd + (-1)^{X}\dd \imath_{X}$. The Euler vector field is degree 0, hence the expression given.}
\begin{equation}
    n\omega = \mathcal{L}_{\varepsilon}\omega = \imath_{\varepsilon}\dd\omega + \dd (\imath_{\varepsilon}\omega) = \dd(\imath_{\varepsilon}\omega)
\end{equation}
where we have used $\dd\omega = 0$. This implies we can take
\begin{equation}
    \vartheta = \tfrac{1}{n}\imath_{\varepsilon}\omega = (\deg Z^{A}) Z^{A}\omega_{AB}\dd Z^{B}
\end{equation}

\subsubsection*{Transgressed QP structure on $\Mcal^{\Xcal}$}

Let $(\Xcal=T[1]X,\dd)$ be a DG manifold with homogeneous coordinates $\sigma, \dd\sigma$. A point $f \in \Mcal^{\Xcal}$ can be defined by how it pulls back the coordinates on $\Mcal$. We have
\begin{equation}\label{eq:pseudo-coords}
    f^{*}Z^{A} = \bm{Z}^{A}(\sigma,\dd\sigma) = Z^{A}_{0}(\sigma) + Z^{A}_{1\,\alpha}(\sigma)\dd\sigma^{\alpha} + ... + \tfrac{1}{d!}Z^{A}_{d\,\alpha_{1}...\alpha_{d}} (\sigma) \dd\sigma^{\alpha_{1}}...\dd\sigma^{\alpha_{d}}
\end{equation}
We use the shorthand $Z^{A}_{k} = \tfrac{1}{k!}Z^{A}_{k\, \alpha_{1}...\alpha_{k}}(\sigma) \dd\sigma^{\alpha_{1}}...\dd\sigma^{\alpha_{k}}$, where $Z^{A}_{k\,\alpha_{1}...\alpha_{k}}(\sigma)$ is a function of degree $\deg Z^{A} - k$. These act as coordinates on $\Mcal^{\Xcal}$. Our conventions are always that the form components come to the right of the function. So, e.g.
\begin{equation}
    Z^{A}_{1} = Z^{A}_{1\, \alpha}(\sigma) \dd \sigma^{\alpha} = (-1)^{A-1}\dd\sigma^{\alpha} Z^{A}_{1\,\alpha}(\sigma)
\end{equation}

We can always define an evaluation map
\begin{equation}
    \arraycolsep = 3pt
    \begin{array}{rcl}
        \mathrm{ev}:\Mcal^{\Xcal}\times \Xcal & \longrightarrow & \Mcal  \\
        (f,\sigma,\dd\sigma) & \longmapsto & f(\sigma,\dd\sigma)
    \end{array}
\end{equation}
We also have the chain map defined by
\begin{equation}
    \arraycolsep = 3pt
    \begin{array}{rcl}
        \mu_{*}: \Omega^{\bullet}(\Mcal^{\Xcal}\times \Xcal) & \longrightarrow & \Omega^{\bullet}(\Mcal^{\Xcal})  \\
        \alpha & \longmapsto & \int_{\Xcal} \alpha
    \end{array}
\end{equation}
The combination $\mu_{*}\mathrm{ev}^{*}:\Omega^{\bullet}(\Mcal)\to \Omega^{\bullet}(\Mcal^{\Xcal})$ is called the transgression map. The QP structure on the mapping space is defined by
\begin{equation}
    \omega_{\Mcal^{\Xcal}} = \mu_{*}\mathrm{ev}^{*}\omega_{\Mcal} \qquad \Theta_{\Mcal^{\Xcal}} = (-1)^{d}\mu_{*}\mathrm{ev}^{*} \Theta_{\Mcal} + (-1)^{n+d+1}\imath_{\dd}\mu_{*}\mathrm{ev}^{*} \vartheta
\end{equation}
where we use the same symbol $\dd$ for the lift of the vector field on $\Xcal$ to $\Mcal^{\Xcal}$.

This can be given more explicitly in the coordinates \eqref{eq:pseudo-coords}. We will use the bold face notation to denote a function, differential form, or coordinate on $\Mcal$ that is pulled back to $\Xcal$ via some function $f\in \Mcal^{\Xcal}$. That is, we effectively take $\bm{f} = \mathrm{ev}^{*} f$. We can then write
\begin{equation}
    \omega_{\Mcal^{\Xcal}} = \tfrac{1}{2}\int_{\Xcal} \delta \bm{Z}^{A} (\omega_{\Mcal})_{AB} \, \delta \bm{Z}^{B}
\end{equation}
Our conventions for integrals is that constants are pulled out from the left. The symplectic form above gives rise to a Poisson bracket which takes the following form on homogeneous functionals $F,G$
\begin{equation}\label{eq:transgressed_PB_full}
    [F,G] = \int_{\Xcal} (-1)^{F} \frac{\delta^{R} F}{\delta \bm{Z}^{A}} (\omega_{\Mcal})^{AB} \frac{\delta G}{\delta \bm{Z}^{B}}
\end{equation}
where
\begin{equation}
    \delta F = \int_{\Xcal}\delta \bm{Z}^{A} \frac{\delta F}{\delta \bm{Z}^{A}} = \int_{X} \frac{\delta^{R} F}{\delta \bm{Z}^{A}}\delta \bm{Z}^{A}
\end{equation}
We can define a functional $F$ via some pulled-back function $\bm{f}$ by
\begin{equation}
    F(\epsilon) = \int_{\Xcal} \bm{f} \epsilon \qquad \forall \, \epsilon \in C^{\infty}(\Xcal)
\end{equation}
Then the Poisson bracket \eqref{eq:transgressed_PB_full} can be expressed nicely as
\begin{equation}
    \left[ \int_{\Xcal} \bm{f} \epsilon, \int_{\Xcal} \bm{g} \eta \right] = (-1)^{(f+n)\epsilon+d}\int_{\Xcal} \bm{(f,g)_{\Mcal}} \epsilon \eta
\end{equation}
where $\bm{(f,g)_{\Mcal}} = \mathrm{ev}^{*}(f,g)_{\Mcal}$.

We can use this to calculate the Poisson bracket on two harmonic generators $Z^{A}_{k}, Z^{B}_{k'}$. Let $e_{k,a}$ be a basis of harmonic $k$-forms and $\tilde{e}_{d-k}^{b}$ be a dual basis of harmonic $d-k$-forms. So
\begin{equation}
    \delta^{b}{}_{a} = \int_{X} e_{k,a}\wedge \tilde{e}^{b}_{d-k}
\end{equation}
Noting that $\Omega^{\bullet}(\Xcal) \simeq C^{\infty}(T[1]X) = C^{\infty}(\Xcal)$, and by expanding $Z^{A}_{k} = Z^{A,a}_{k}e_{k,a}$ with $Z^{A,a}_{k}$ some constant coefficient, we have
\begin{equation}
    Z^{A,a}_{k} = \int_{\Xcal} Z^{A}_{k} \tilde{e}^{a}_{d-k} = \int_{\Xcal} \bm{Z}^{A} \tilde{e}^{a}_{d-k}
\end{equation}
We then see that we get an induced Poisson bracket on the coefficients given by
\begin{equation}
\begin{split}
    \left[Z^{A,a}_{k},Z^{B,b}_{k'}\right] &\equiv \left[\int_{\Xcal} \bm{Z}^{A} \tilde{e}^{a}_{d-k}, \int_{\Xcal} \bm{Z}^{B} \tilde{e}^{b}_{d-k'} \right] \\
    &= (-1)^{(A+n)(d-k)+d} \int_{\Xcal} \bm{(Z^{A},Z^{B})_{\Mcal}} \tilde{e}^{a}_{d-k} \wedge \tilde{e}^{b}_{d-k'} \\
    &= (-1)^{(A+n)(d-k)+d}(-1)^{A}\omega^{AB} \int_{\Xcal} \tilde{e}^{a}_{d-k} \wedge \tilde{e}^{b}_{d-k'} \\
    &= (-1)^{(A+n)(d-k)+d}(-1)^{A}\omega^{AB} \kappa^{ab}\delta_{k+k',d}
\end{split}
\end{equation}
where we have assumed Darboux coordinates, so the $\omega^{AB}$ are constant, and where
\begin{equation}
    \kappa^{ab} = \int_{X} \tilde{e}^{a}_{d-k}\wedge \tilde{e}^{b}_{k}
\end{equation}
We use this to find the symplectic form of the reduced theory.

\section{Coisotropic reduction of graded Poisson algebras}
\label{app:CoisoreductionOfPoisson}

Let $\mathcal P$ be a graded algebra with a graded Poisson bracket $[\bullet,\bullet]$ of degree $-P$ along with a left derivation $\mathcal V$ of $\mathcal P$, possibly \emph{hamiltonian} (i.e.~given by Poisson brackets, so $\mathcal V=[\mathcal H_{\mathcal V},\bullet]$ for $\mathcal H_{\mathcal V}\in \mathcal P$). We will explain how all of these objects behave under \emph{coisotropic reduction}. The derivation is as in the ungraded case considered originally by Weinstein and \'Sniatycki \cite{sniatycki1983reduction}.

If $\mathcal I$ is a (multiplicative, degree-homogeneous) ideal of $\mathcal P$, it is a \emph{coisotrope} if it is a Poisson subalgebra, i.e.~$[\mathcal I,\mathcal I]\subseteq \mathcal I$. Then the \emph{coisotropic reduction} of $\mathcal P$ with respect to $\mathcal I$ is the quotient
\be
\overline{\mathcal P}\equiv N(\mathcal I)/\mathcal I
\ee
where $N(\mathcal I)\equiv\{ f\in \mathcal P| [f,\mathcal I]\subseteq\mathcal I\}$ is the Poisson normaliser of $\mathcal I$. Then the bracket on $\overline{\mathcal P}$ is defined in terms of the bracket $[\bullet,\bullet]$ via
\be
[\Pi f,\Pi g]_{\bar{\mathcal P}}\equiv \Pi{[f,g]}\,.
\ee
where $\Pi f$ is the equivalence class $f+\mathcal I$. For any $\mathcal P$ derivation $\mathcal V$ we define its reduction $\overline{\mathcal V}$ via
\be
\overline{\mathcal V}(\Pi{f})=\Pi{\mathcal V(f)}\qquad f\in \mathcal P\,.
\ee

\begin{theorem}\label{thm:coisoreduction}
Given any coisotrope $\mathcal I$, the bracket $[\bullet,\bullet]_{\overline{\mathcal P}}$ is well-defined. It is moreover a Poisson bracket of degree $-P$, and so $\overline{\mathcal P}$ is a graded Poisson algebra.

If the derivation $\mathcal V$ on $\mathcal P$ preserves the Poisson structure ($\mathcal V[f,g]=[\mathcal Vf,g]\pm[f,\mathcal V g]$) and the coisotrope ($\mathcal V(\mathcal I)\subseteq \mathcal I$) then the reduced derivation $\overline{\mathcal V}$ is well-defined.

Finally if $\mathcal V$ is furthermore hamiltonian with hamiltonian $\mathcal H_{\mathcal V}\in\mathcal P$ (so $\mathcal V=[\mathcal H_{\mathcal V},\bullet]$) then $\overline{\mathcal V}$ is hamiltonian with hamiltonian $\Pi(\mathcal H_{\mathcal V})$. (In this latter case $\mathcal V$ automatically preserves the Poisson structure, but  the condition $\mathcal V(\mathcal I)\subseteq \mathcal I$ implies $[\mathcal H_{\mathcal V},\mathcal I]\subseteq \mathcal I$.)
\end{theorem}
If all derivations we are interested in are in fact hamiltonian (which is the case in the main text) then we just need to check that the ideal $\mathcal I$ is a coisotrope and that $[\mathcal H_{\mathcal V},\mathcal I]\subseteq\mathcal I$.
\begin{proof}
The bracket $[\bullet,\bullet]_{\overline{\mathcal P}}$ is well-defined because
\be
[\Pi f,\Pi g]_{\overline{\mathcal P}}=\Pi{[f+\mathcal I,g+\mathcal I]}= \Pi([f,g]+[f,\mathcal I]+[\mathcal I,g]+[\mathcal I,\mathcal I])=\Pi{[f,g]}
\ee
where the last three terms in the second equality vanish because $f,g\in N(\mathcal I)$ and $[\mathcal I,\mathcal I]\subseteq \mathcal I$. This new bracket inherits the antisymmetry and Jacobi identity properties from $[\bullet,\bullet]$. Since furthermore $\mathcal I$ is homogeneous in degree, $\Pi{f}$ will have a well-defined degree, and so the new bracket defines a graded Poisson algebra structure.

Similarly since $\mathcal V(f+\mathcal I)=\mathcal V(f)+\mathcal V(\mathcal I)$ we have that $\overline{\mathcal V}$ is well-defined on $\mathcal P/\mathcal I$ when $\mathcal V(\mathcal I)\subseteq \mathcal I$. We then need to show that it preserves the subspace $N(\mathcal I)/\mathcal I=\overline{\mathcal P}$. Since $\mathcal V$ preserves the Poisson bracket we have
\be
[\mathcal V f,\mathcal I]= \mathcal V[f,\mathcal I]\pm [f,\mathcal V\mathcal I]
\ee
If $f\in N(\mathcal I)$ this becomes $\mathcal V(\mathcal I) \pm [f,\mathcal V\mathcal I]$ which lies in the coisotrope when $\mathcal V(\mathcal I)\subset \mathcal I$. Therefore $\mathcal V(f)$ lies in $N(\mathcal I)$.

Finally if $\mathcal V=[\mathcal H_{\mathcal V},\bullet]$ then $\Pi{\mathcal V}(\Pi{f})=\Pi{\mathcal V(f)}=\Pi{[\mathcal H_{\mathcal V},f]}=[\Pi{\mathcal H_{\mathcal V}},\Pi{f}]_{\Pi{\mathcal P}}$,
which completes the proof. 
\end{proof}

\bibliographystyle{JHEP}
\bibliography{citations}

\end{document}